\documentstyle[epsfig,epsf]{mn}

\newif\ifAMStwofonts

\ifoldfss
  \ifCUPmtlplainloaded \else
    \NewTextAlphabet{textbfit} {cmbxti10} {}
    \NewTextAlphabet{textbfss} {cmssbx10} {}
    \NewMathAlphabet{mathbfit} {cmbxti10} {} 
    \NewMathAlphabet{mathbfss} {cmssbx10} {} 
  \fi
  \ifAMStwofonts
    \ifCUPmtlplainloaded \else
      \NewSymbolFont{upmath} {eurm10}
      \NewSymbolFont{AMSa} {msam10}
      \NewMathSymbol{\upi}     {0}{upmath}{19}
      \NewMathSymbol{\umu}     {0}{upmath}{16}
      \NewMathSymbol{\upartial}{0}{upmath}{40}
      \NewMathSymbol{\leqslant}{3}{AMSa}{36}
      \NewMathSymbol{\geqslant}{3}{AMSa}{3E}

      \let\leq=\leqslant 
      \let\geq=\geqslant 
    \fi
  \fi
\fi 

\ifnfssone
  \newmathalphabet{\mathit}
  \addtoversion{normal}{\mathit}{cmr}{m}{it}
  \addtoversion{bold}{\mathit}{cmr}{bx}{it}
  \newmathalphabet{\mathbfit} 
  \addtoversion{normal}{\mathbfit}{cmr}{bx}{it}
  \addtoversion{bold}{\mathbfit}{cmr}{bx}{it}
  \newmathalphabet{\mathbfss} 
  \addtoversion{normal}{\mathbfss}{cmss}{bx}{n}
  \addtoversion{bold}{\mathbfss}{cmss}{bx}{n}
  \ifAMStwofonts
    \ifCUPmtlplainloaded \else
      \UseAMStwoboldmath
      \makeatletter
      \new@mathgroup\upmath@group
      \define@mathgroup\mv@normal\upmath@group{eur}{m}{n}
      \define@mathgroup\mv@bold\upmath@group{eur}{b}{n}
      \edef\UPM{\hexnumber\upmath@group}
      \new@mathgroup\amsa@group
      \define@mathgroup\mv@normal\amsa@group{msa}{m}{n}
      \define@mathgroup\mv@bold\amsa@group{msa}{m}{n}
      \edef\AMSa{\hexnumber\amsa@group}
      \makeatother
      \mathchardef\upi="0\UPM19
      \mathchardef\umu="0\UPM16
      \mathchardef\upartial="0\UPM40
      \mathchardef\leqslant="3\AMSa36
      \mathchardef\geqslant="3\AMSa3E

      \let\leq=\leqslant 
      \let\geq=\geqslant 
    \fi
  \fi
\fi 

\ifnfsstwo
  \DeclareMathAlphabet{\mathbfit}{OT1}{cmr}{bx}{it}
  \SetMathAlphabet\mathbfit{bold}{OT1}{cmr}{bx}{it}
  \DeclareMathAlphabet{\mathbfss}{OT1}{cmss}{bx}{n}
  \SetMathAlphabet\mathbfss{bold}{OT1}{cmss}{bx}{n}
  \ifAMStwofonts
    \ifCUPmtlplainloaded \else
      \DeclareSymbolFont{UPM}{U}{eur}{m}{n}
      \SetSymbolFont{UPM}{bold}{U}{eur}{b}{n}
      \DeclareSymbolFont{AMSa}{U}{msa}{m}{n}
      \DeclareMathSymbol{\upi}{0}{UPM}{"19}
      \DeclareMathSymbol{\umu}{0}{UPM}{"16}
      \DeclareMathSymbol{\upartial}{0}{UPM}{"40}
      \DeclareMathSymbol{\leqslant}{3}{AMSa}{"36}
      \DeclareMathSymbol{\geqslant}{3}{AMSa}{"3E}

      \let\leq=\leqslant 
      \let\geq=\geqslant 
    \fi
  \fi
\fi 

\ifCUPmtlplainloaded \else
  \ifAMStwofonts \else 
    \def\upi{\pi}
    \def\umu{\mu}
    \def\upartial{\partial}
  \fi
\fi

\title{Study of Variable Stars in the MOA Database\footnotemark: \\ Long-Period Red Variables in the Large Magellanic Cloud\\ \vspace{0.5cm}\Large{$^{\star}$ Partly based on DENIS data obtained at the European Southern Observatory}}

\author[S. Noda et al.]{
    S.~Noda,$^{1}$ M.~Takeuti,$^{2}$ F.~Abe,$^{1}$ I.A.~Bond,$^{3}$ R.J.~Dodd,$^{3,4,14}$
\newauthor    J.B.~Hearnshaw,$^{5}$ M.~Honda,$^{6}$ M.~Honma,$^{7}$ J.~Jugaku,$^{8}$ S.~Kabe,$^{9}$ Y.~Kan-ya,$^{10}$
\newauthor    Y.~Kato,$^{1}$ P.M.~Kilmartin,$^{3, 5}$ Y.~Matsubara,$^{1}$ K.~Masuda,$^{1}$ Y.~Muraki,$^{1}$ 
\newauthor    T.~Nakamura,$^{11}$ G.R.~Nankivell,$^{12}$ C.~Noguchi,$^{1}$ K.~Ohnishi,$^{13}$ M.~Reid,$^{14}$
\newauthor    N.J.~Rattenbury,$^{3}$ To.~Saito,$^{15}$ H.~Sato,$^{11}$ M.~Sekiguchi,$^{6}$ J.~Skuljan,$^{5}$ 
\newauthor    D.J.~Sullivan,$^{14}$ T.~Sumi,$^{1}$ Y.~Watase,$^{9}$ S.~Wilkinson,$^{14}$ R.~Yamada,$^{1}$ 
\newauthor    T.~Yanagisawa,$^{1}$ P.C.M.~Yock,$^{3}$ and M.Yoshizawa,$^{16}$
\\
$^1$Solar-Terrestrial Environment Laboratory, Nagoya University, Nagoya 464-8601, Japan\\
$^2$Astronomical Institute, Tohoku University, Sendai 980-8578, Japan\\
$^3$Dept. of Physics, University of Auckland, Auckland, N.Z.\\
$^4$Carter National Observatory, Wellington, N.Z.\\
$^5$Dept. of Physics and Astronomy, University of Canterbury, Christchurch, N.Z.\\
$^6$Institute for Cosmic Ray Research, University of Tokyo, Kashiwa, 277-8582, Japan\\
$^7$VERA Project Office, National Astronomical Observatory, 2-21-1 Osawa, Mitaka, Tokyo 181-8588, Japan\\
$^8$Institute for Civilization, Tokai University, Japan\\
$^9$High Energy Accelerator Research Organization (KEK), Tsukuba 305-0801, Japan\\
$^{10}$Institute of Astronomy, University of Tokyo, Tokyo 181-0015, Japan\\
$^{11}$Research Inst. Fundamental Physics, Kyoto University, Kyoto 606-8502, Japan\\
$^{12}$Lower Hutt, N.Z.\\
$^{13}$Nagano National College of Technology 381-8550, Japan\\
$^{14}$School of Chemical and Physical Sciences, Victoria University, Wellington, N.Z\\
$^{15}$Tokyo Metropolitan College of Aeronautics, Tokyo 140-0011, Japan\\
$^{16}$National Astronomical Observatory, Mitaka, Tokyo 181-8565, Japan\\
}

\date{Accepted 0000 January 00.
      Received 0000 December 00;
      in original form }

\pagerange{\pageref{firstpage}--\pageref{lastpage}}
\pubyear{1994}

\begin{document}

\maketitle

\label{firstpage}

\begin{abstract}
One hundred and forty six long-period red variable stars in the Large
Magellanic Cloud (LMC) from the three year MOA project database were analysed.
A careful periodic analysis was performed on these stars and a catalogue of
their magnitudes, colours, periods and amplitudes is presented.
We convert our blue and red magnitudes to $K$ band values using 19 oxygen-rich
stars. A group of red short-period stars separated from the Mira sequence has
been found on a ($\log P, K$) diagram. They are located at the short period
side of the Mira sequence consistent with the work of Wood and Sebo (1996).
There are two interpretations for such stars; a difference in pulsation mode
or a difference in chemical composition. We investigated the properties of
these stars together with their colour, amplitude and periodicity. We conclude
that they have small amplitudes and less regular variability. They are likely to be
higher mode pulsators. A large scatter has been also found on the long period
side of the ($\log P, K$) diagram. This is possibly a systematic
spread given that the blue band of our photometric system covers both standard
$B$ and $V$ bands and affects carbon-rich stars.
\end{abstract}


\section{Introduction}

There are many types of variable stars whose mechanism for variability
(mainly pulsation, rotation, eruptive processes, and geometric
effects) is well studied and well understood observationally and
theoretically.  Among them, the pulsating variable stars have been
investigated vigorously because they yield information on their
stellar mass and their interior physical properties. 
They are a powerful tool for measuring distances in the Galaxy and
extra-galactic systems (1) because they are luminous and (2) because
they have been shown empirically to obey a period-luminosity (PL) relation.

The PL relation for classical Cepheids was
discovered by Leavitt (1908) and used to measure the distance of
galaxies. However, there is a discrepancy between the distance to the
LMC derived from the PL relations for Cepheids and that derived for RR
Lyraes (Walker 1992). This discrepancy may be due to uncertainties in
the absolute calibration and a sensitivity to metallicity effects
(Caputo 1997). It is important to establish PL relations of any new
objects which are independent of these variable stars.

A PL relation for long period red variables (Miras and semi-regular
variables (hereafter SRs)) will be a useful distance indicator because these stars are
numerous and have high luminosities. However, they are the least
studied star type because of their complex pulsation mechanism.
Difficulties arise from uncertainties in our understanding of the
convective energy transport and the opacity resulting from the many
molecular absorption lines.

We present in this paper the results of our study of LMC Miras and SRs
using the MOA database. This database was generated from the MOA
project microlensing survey.

A PL relation for Miras of the LMC was presented by Glass \&
Lloyd-Evans
(1981) and Glass \& Feast (1982). They observed 11 Miras in the
infrared bands $J, H, K$. However, there was a large amount of scatter
in the plot ($\sigma\sim 0.25$ mag) because their observation time was
limited and it did not cover the whole phase of variations.

To reduce the scattering in the plot due to poor phase coverage,
extensive infrared observations have been made (Feast et al. 1989;
Hughes \& Wood 1990), and a narrow PL relation was obtained ($\sigma
\sim 0.13$ mag). Feast (1996) showed that the Miras in the LMC follow
a narrow PL relation in both absolute bolometric magnitude, $M_{bol}$,
and absolute $K$-magnitude, $M_{K}$, and he concluded that all Miras
pulsate in the same mode.

Wood \& Sebo (1996) found two sequences in a $K$ band PL diagram for the
small amplitude SRs in the LMC. One was the well-known Mira sequence,
and the other (eight stars) was located at the position which is
parallel to the Mira sequence and their periods were approximately
half the periods of Miras on the PL diagram.  This discovery
encouraged a hypothesis that the pulsation mode of these stars in the
secondary sequence must be in a higher mode.
Using the Hipparcos database of nearby stars, Bedding \& Zijlstra
(1998) discovered the SRs which are located in the secondary sequence
of Wood \& Sebo (1996). 
Bedding and Zijlstra interpreted the sequence of SRs as
evolutionary (see Whitelock (1986)).

Feast (1999) suggested that both Miras and SRs
pulsate in the same mode, possibly both in the first overtone. 
The arguments were based on the angular diameter measurements of nearby
Miras and the PL relationship of the Miras and SRs in globular clusters. 
He agreed with Bedding and Zijlstra (1998) that
the sequence which is located on the
shorter period side of the Mira sequence in the PL plane is an
evolutionary track ending at the Mira sequence. Feast also suggested
that the Mira sequence would be a collection of variable stars with
different chemical abundances and/or with different stellar mass. In
this case, it was not necessary that Miras and SRs pulsate in
different modes.

As described above, there are two theories on the short-period stars
in the PL diagram for long-period red variable stars. Considering that
the pulsation features of the Miras and SRs depend on several factors,
other physical properties must also be considered. The study of
stellar colour will be interesting because this gives us the
temperature of stellar surface, stellar radius and mass which are the
important factors of the pulsation period.

Recently, the continuous observation of a great number of stars using
CCD cameras which cover a large image area became possible.  The
recent report of Wood (2000) based on the $K$ band photometry and the
period derived from the MACHO database indicated plausible sequences
on the PL diagram. This diagram shows at least five sequences falling
parallel to the Mira sequence. According to Wood,
the three sequences containing the Mira sequence were explained as
being different pulsation modes.  However, more information is
required to settle this issue. The features of the other two sequences
are unknown, and it is expected that further independent observations and detailed investigation will reveal the nature of these stars.

We worked on the problem using the MOA database which contained data
for a large number of stars in the LMC. The period, magnitude colour, amplitude and periodicity for each star was determined.

We found an independent group of short period stars in a sample of 146 stars. We investigated the properties of these stars through their colour, amplitude and periodicity, and discuss the possibility of that they are higher mode pulsators.

Following an introduction to the MOA project, we describe our
observation and photometry systems in Sections 2 and 3. To obtain the light
curves of variable stars, a careful calibration of each image frame is required. This procedure is described in Section 4. In
Sections 5 and 6, the data reduction and period analysis methods are
described. In Section 7, the colour-magnitude diagram (CMD) of the LMC
stars observed using our system is presented and the period-magnitude
relation will be given in Section 8. 
The period-colour relation is presented and discussed
in Section 9.
The result of the conversion of
MOA magnitudes to $K$ band values will be given in Section 10, along
with a $K, \log P$ diagram. In Section 11 we discuss the possibility
of the existence of multiple sequences in the ($\log P, K$) plane with
the amplitude distribution. In Section 12 we summarise our results.

\section{The MOA Project Observations}

\subsection{Observational Instruments}
The MOA project commenced observations in May 1996 with the aim of
observing microlensing events towards the LMC, SMC and Galactic
Bulge. The observational program was undertaken at the University of
Canterbury's Mt. John Observatory ($44^{\circ} S, 1030 \rm m$ above sea
level) in the centre of the South Island of New Zealand.

The MOA project uses a 61cm Ritchey-Chr\'{e}tien Cassegrain telescope
with modified optics to give a wide FOV at f/6.25.

For the first set of observations in 1996 (Series 1), we used the
original f/13.5 telescope optics and monitored $\sim 3\times 10^5$
stars in three LMC fields with the mosaic CCD camera,
MOA-cam1. MOA-cam1 was constructed with nine TI TC215 CCDs with
$1000\times 1018$ pixels whose pixel size was $12 \mu$m $\times 12 \mu$m. 
From January 1997 to July 1998, in Series 2, the optics were
altered to f/6.25. MOA-cam1 covered 3 fields in the LMC bar ($1\times
10^6$ stars) and 2 fields in the SMC ($\sim 4\times 10^5$ stars).  The
current set of observations, Series 3, started in August 1998. For
this series, a new camera, MOA-cam2 is in use. MOA-cam2 consists of
three abutted SITe $2047\times 4095$ pixel thinned, back-illuminated
CCDs with $15\mu$m $\times 15 \mu$m pixels. The pixel size is
$0.81''$, giving a field of view of $0.5^{\circ}\times 0.9^{\circ}$
for each chip. The details of MOA-cam2 are in Yanagisawa et
al. (2000). In this series we are observing 16 LMC fields ($4.4\times
10^6$ stars), 8 SMC fields ($9.3\times 10^5$ stars) and 14 Galactic
Bulge fields.

We used two broad-band filters to maximise photon collection. The MOA
blue filter ($MOA_B$) covers $395-620 n$m ($\lambda_{\rm eff} \sim 500$ nm)
and MOA red filter ($MOA_R$) covers $620-1050$ nm ($\lambda_{\rm eff}
\sim 700$ nm).  The spectral transmission curves for each filter are
shown in Fig.~1.

\begin{figure}
  \centerline{\psfig{figure=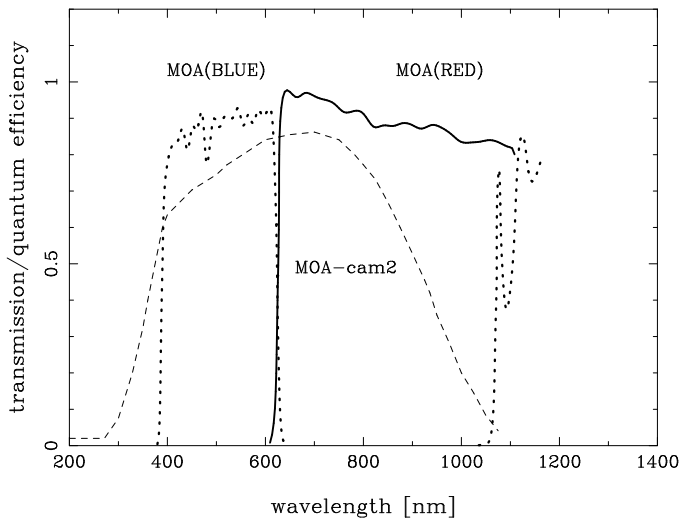,width=9cm}}
  \caption{The transmission of the filters and quantum efficiency of MOA-cam2 
	   are shown as a function of the wavelength. We used two broad-band 
	   filters ($ MOA_B$, $MOA_R$), where $MOA_B$ covers standard B and V, 
	   and $MOA_R$ covers the Kron-Cousins $RI$ region.}
\label{fig1}
\end{figure}

The MOA instrumental photometry has been transformed to the standard
Johnson $V$ and Kron-Cousins $R$ ($R_{KC}$) by comparisons with about one
hundred stars imaged by Hubble Space Telescope (HST). We describe
these details in Section 4.

The variable star study based on the MOA database was reported in
Hearnshaw, Bond, Rattenbury, et al. (2000), Noda, Takeuti, Bond, et
al. (2000), and Takeuti, Noda, Bond, et al. (2000).

\subsection{OBSERVATION DETAILS}

Regular 300 second exposures were made towards the Large Magellanic
Cloud and the Small Magellanic Cloud. 180 second exposures were made
for the Galactic Bulge (GB) fields. Under the current system, it takes
70 seconds to read out the 3 CCD images for each exposure. Dark frames
for each exposure were prepared every photometric night and dome flats
or dark-sky flats were also frequently taken.

Observational data were archived on DLT tapes and all the data were
analysed in both Japan and New Zealand. Series 1, 2
and 3 images were analysed using DoPHOT photometry software
(Schechter et al. 1993). The size of the Series 2 database created by
the subsequent photometry is $\simeq$ 40 Gbytes.

A new online analysis system based on image subtraction has been
developed in New Zealand. In the year 2000 this system was able to
find several microlensing events in real time and issue alerts to the
astronomical community via the WWW\footnote{http://www.phys.canterbury.ac.nz/${\sim}$physib/alert/alert.html}. We are re-analyzing some parts of our data with this method.

\section{DATA ANALYSIS}

Before undertaking data analysis using the DoPHOT program, all image
files were prepared with dark and the flat field frames. DoPHOT was
used in a fixed position warm start mode, using the best photometric
images as templates.  It takes about 50 minutes to analyze one LMC
frame with our computer.  There were more than 400 
measurements in the Series 2 (1.6 year) and Series 3 (1.4 year) database. 
We obtain only relative magnitudes from the DoPHOT analysis. We also use
non-standard filters, thus a transformation of our instrumental
magnitude to the standard $V$ and $R_{KC}$ is required.  This
transformation is described in the next section.

\section{CALIBRATION FOR MOA PHOTOMETRY}

The transformation of the MOA photometry to the standard system was
carried out using Hubble Space Telescope (HST) image data. More
details are given by Kato (2000). We defined the transform relation
for MOA-cam1 and MOA-cam2 separately because their CCD quantum
efficiencies were quite different (the quantum efficiency of MOA-cam2
was twice as high as that for MOA-cam1).  Although the filter system
is exactly the same, this will cause serious differences in the
relative magnitudes returned by DoPHOT. The DoPHOT magnitudes were
calculated using photon quantities only, so a difference in CCD
quantum efficiency will affect the DoPHOT output directly. To get
standard magnitudes for MOA-cam2, we performed the procedure described below.

Photons coming
from a star show a complicated scattering pattern around a central
position caused by atmospheric turbulence. A point-spread function
(PSF) models the scatter. The DoPHOT code estimated each magnitude
using an elliptical Gaussian with seven free parameters as a model for
a stellar PSF. A common PSF was adopted for all stars (bright stars and
faint stars) in a frame, however it was clear that there were
significant differences between the profiles for bright stars and that
for faint stars. The former were weighted by the photon statistics of
the star itself, and the latter were weighted by the photon statistics
of the sky, so the latter should show larger scattering. The detailed
study of the PSF indicated that the function often differs from the
elliptical Gaussian (Alard \& Lupton 1998). The error due to this
discrepancy is approximately $\pm$0.05 mag for the bright stars and
$\pm$0.09 mag for the faint stars. It was not a critical problem if we
define a mean magnitude with more than 100 measurement points to study
stellar intensity, because the scatter due to PSF model inadequacy
usually occurs at random.

As we were using large CCD chips, we have to determine the position
dependent error for each CCD chip. In order to investigate this, we
observed the globular cluster NGC3201 at various positions on each
chip, and compared stellar magnitudes with published magnitudes
obtained using $B,V,R,I$ filters by Alcaino et al. (1989). We found that
the error was smaller ($\pm 0.015$ mag) than the DoPHOT PSF photometry
error ($\leq \pm 0.09$ mag). This was a systematic error however, and
did not disappear even if we deal with a mean magnitude.

In the third step, also using the NGC3201 data, we computed the chip
to chip differences for the magnitude zero point:

\begin{eqnarray}
    MOA_{B(chip3)} = MOA_{B(chip1)} - 0.23\\
    MOA_{R(chip3)} = MOA_{R(chip1)} - 0.31\\
    MOA_{B(chip3)} = MOA_{B(chip2)} - 0.34\\
    MOA_{R(chip3)} = MOA_{R(chip2)} - 0.41
\label{eq:3chips}
\end{eqnarray}

We used the chip3 scale as our standard as it gave the most linear response.


For the final step, we converted our $MOA_B$ and $MOA_R$ magnitudes to
standard Johnson $V$ and Kron-Cousins $R$ ($R_{KC}$) respectively. The
brightness of stars was acquired from an ADU (Analog Digital Unit, 1
ADU corresponded to 1.3 electrons for MOA-cam2) value which
corresponded to the number of photons measured with the assumed PSF
via the DoPHOT photometry. We define the zero-point of magnitude for
MOA as 1 ADU. Therefore all of the obtained magnitudes are negative,
and we have to correct the zero-point of our magnitude in order to
obtain actual magnitudes. 
We analyzed the HST data of the LMC obtained with the Wide Field Planetary
Camera2 (WFPC2) to correct for atmospheric seeing. Since the filters for WFPC2
had been transformed to the Johnson $V$ and $R_{KC}$ by Holtzman et al. (1995), we
were able to compare the converted HST magnitudes directly with those of
our one hundred stars.

Finally, we found a relation to convert the $V_m$ and $R_m$ for MOA-cam2;

\vspace{10pt}
\begin{equation}
V_m = MOA_B - 0.16(\pm0.01) *(MOA_B - MOA_R) + 27.41(\pm0.01)
\label{calib_b2}
\end{equation}
\begin{equation}
R_m = MOA_R + 0.29(\pm0.01) *(MOA_B - MOA_R) + 27.24(\pm0.01)
\label{eq:calib_r2}
\end{equation}
\begin{equation}
Colour(V_m - R_m) = 0.53 * Colour(MOA_B - MOA_R) + 0.18
\label{eq:calib_br2}
\end{equation}

For MOA-cam1;

\begin{equation}
V_m = MOA_B - 0.10(\pm0.01) *(MOA_B - MOA_R) + 25.55(\pm0.01)
\label{calib_b1}
\end{equation}
\begin{equation}
R_m = MOA_B + 0.31(\pm0.02) *(MOA_B - MOA_R) + 25.45(\pm0.1)
\label{calib_r1}
\end{equation}
\begin{equation}
Colour(V_m - R_m) = 0.59 * Colour(MOA_B - MOA_R) + 0.10
\label{calib_br1}
\end{equation}

Hereafter, we deal with $V_m$ and $R_m$ as the MOA standard
magnitudes. In Fig.~2, we show the Colour Magnitude Diagram (CMD)
obtained by MOA-cam2 for our observation fields in the LMC bar. The
limiting magnitude was around 20.5 mag for MOA-cam2, and 19 mag for
MOA-cam1. Since our passbands were broad and covered frequencies from
violet to near infra-red bands, the colour is not sensitive for blue stars.

\begin{figure}
  \centerline{\psfig{figure=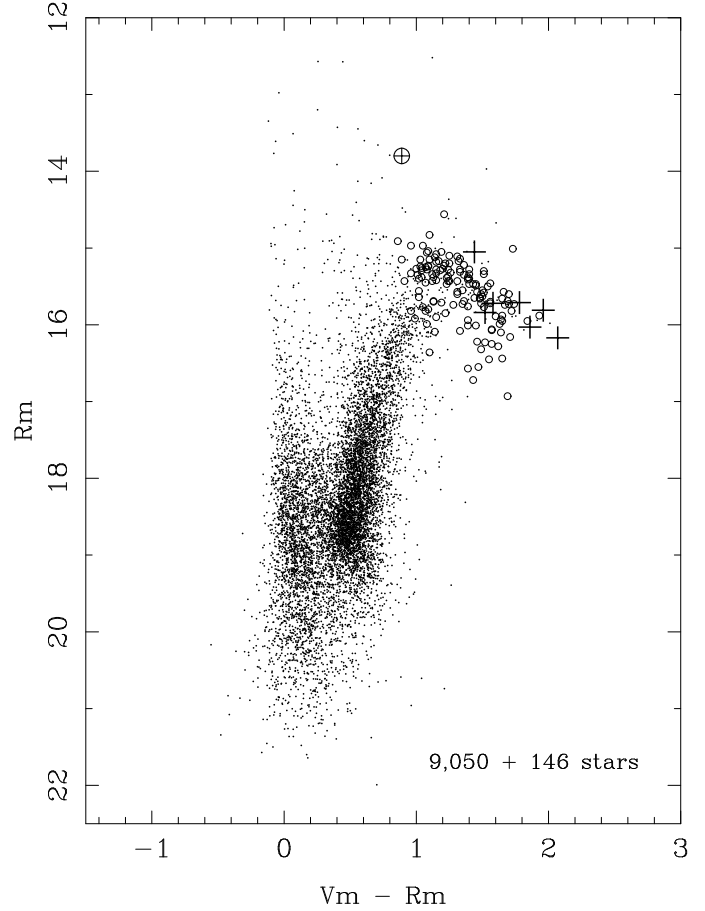,width=9cm}}
  \caption{Colour-Magnitude Diagram (CMD) for the 146 stars in our sample with 
	   9,050 randomly chosen stars (dots). The vertical axis indicates 
	   $R_m$. All 146 stars are located on the tip of Giant Branch, 
	   which means they are in the AGB stage. The seven short period stars are indicated with + symbols. The $\oplus$ mark indicates nlmc3-214266 (see Section 7). The other stars are plotted as open circles.}
\label{fig2}
\end{figure}

To minimize the effects of the non-uniformity of the sky
and system on the raw magnitudes (i.e. DoPHOT output magnitudes)
we:
\begin{enumerate}
\item  Divide the CCD chips into small regions (200 $\times$ 200 pixels)
\item Select eighty bright stars which had more than 100
measurement points in each small region
\item Regard stars with a small scatter as `non-variable' stars, 
i.e., where the number of observed points more than 
4 sigma from the average magnitude was less than 15 
\item Calculate magnitudes for all stars by comparison with the `non-variable' stars. That is, the MOA standard magnitudes were determined relative
to the magnitudes of the near `non-variable' stars in each small region. 
\end{enumerate}

Through the revision of our calibration procedure, we obtained
progressive and interesting diagrams compared with our previous
reports (Hearnshaw, Bond, Rattenbury, et al. 2000; Noda, Takeuti,
Bond, et al. 2000; Takeuti, Noda, Bond, et al. 2000). Details are
given in later sections.

\section{Data Reduction}
We selected long-period red variables from the Series 2 database
using the following criteria:

\begin{enumerate}
\item Each star should be `photometric' and have a `single star shape' as judged by DoPHOT.
\item Light curves should be sufficiently long ($\geq$ 100 points).
\item Light curves of each star must have large variability,
i.e., the reduced $\chi^2$ for a fit to mean brightness should be larger than
$\chi^2_{Vm}\geq\chi^2_0$ and $\chi^2_{Rm}\geq\chi^2_0$.
Here $\geq \chi^2_0$ corresponded to the upper $10\%$ limit of the reduced
$\chi^2$ distribution. For the MOA database $\chi^2_0=4.3$.

To obtain large amplitude variables, we applied the additional criteria:\\

\item Low amplitude stars at an early stage of the study are rejected by removing those stars which had a magnitude difference between the fourth
highest and fourth lowest magnitude values of less than 1.3 magnitudes.
This difference was larger than the amplitude $\delta R_m$ determined
in Section 6.
This selection was done for the results of MOA-cam1 system without using
the colour dependence in equations (8) - (10). 
This sample of large amplitude stars might be affected when we use the correct colour relation.
\item We select redder stars for large
amplitude; $(V_m-R_m)\geq0.4$ for the MOA-cam1 system.
\end{enumerate}


We obtained 1,067 light curves for the candidates of variable stars with large
amplitude after applying criteria (iv) and (v) on the Series 2 database.
Light curves from Series 2 and Series 3 were cross referenced to stars from
their respective templates. We rejected unreliable light curves, and
finally obtained 908 light curves. In Fig.~3 we show an example of a light curve.

We also found 485 small amplitude variable star candidates by applying a further
criterion after numbers (i) to (iii) above. We analyse the large amplitude variables in this work. We will present the small amplitude variable star results in a future paper.

\begin{figure}
  \centerline{\psfig{figure=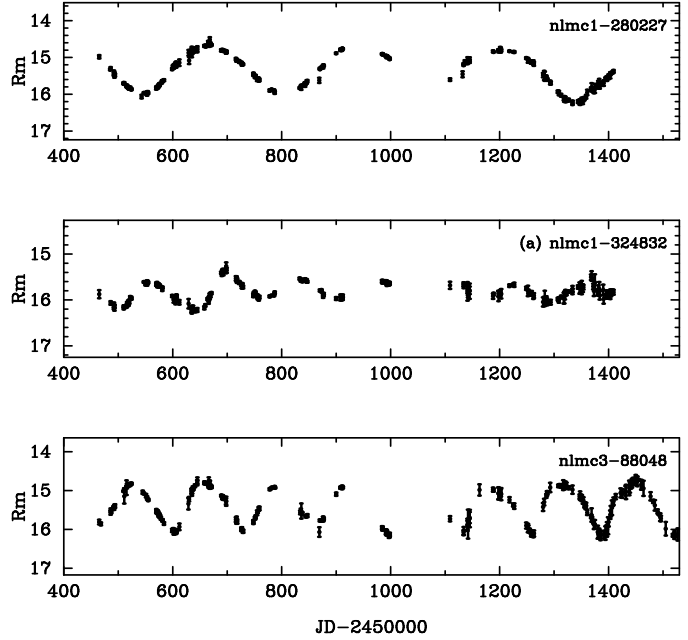,width=9cm}}
   \caption{Examples of light curves for $R_m$.
	    The periods for these stars were identified as 268.0, 133.5, and
	    131.2 days respectively by the PDMM and PDM algorithms (see Section 6).
	    The time axis spans from January 1997 to December 1999.
	    The error of each observation point is the DoPHOT PSF error.
	    The stars nlmc1-324832 and nlmc3-88048 were a short-period
	    red star and a short-period blue star discussed in Section 11.
	    Lack of data after $JD=2451400$ for nlmc1-280227 and nlmc1-324832
	    were due to instrumental problems with chip 1 on MOA-cam2.}
\label{fig3}
\end{figure}

\section{Period Analysis (PDMM)}

As the first stage in the period analysis, we used the {\bf P}hase
{\bf D}ispersion {\bf M}inimization method for {\bf M}OA (PDMM) which is a
PDM method (Stellingwerf 1978) modified to operate on large photometry
datasets. The PDMM algorithm finds a likely light curve period together with a reliability
parameter, $\theta$, for a large number of light curves automatically.
$\theta = 0$ indicates a complete cyclic change and $\theta = 1$ indicates
no periodicity. The amplitude, the difference between the magnitudes of the
brightest and the faintest bins, was also calculated. We converted the
magnitudes of each bin to intensities and calculated the mean intensity. We
denote the mean magnitude values as $\langle V_m \rangle$ and $\langle R_m
\rangle$. We executed this code using both colours ($V_m, R_m$) for the 908
selected light curves.

Since our dataset included scattered data points, the PDMM algorithm 
sometimes gave integral multiples or sub-multiples of the period,
or in the worst case  an incorrect result.
To avoid simple errors in the periods obtained by the automatic analysis,
we checked all code output carefully, and ran the
PDM code manually to determine the most likely periods. We accepted a
period which had good agreement between both colours. In Fig.~4, we
present an example of PDMM figures.  The upper diagram shows the
reliability of the testing period. A small value of $\theta$ indicates
a likely value for the period. The bottom diagram shows the light curve folded
with the most likely period.

\begin{figure}
  \centerline{\psfig{figure=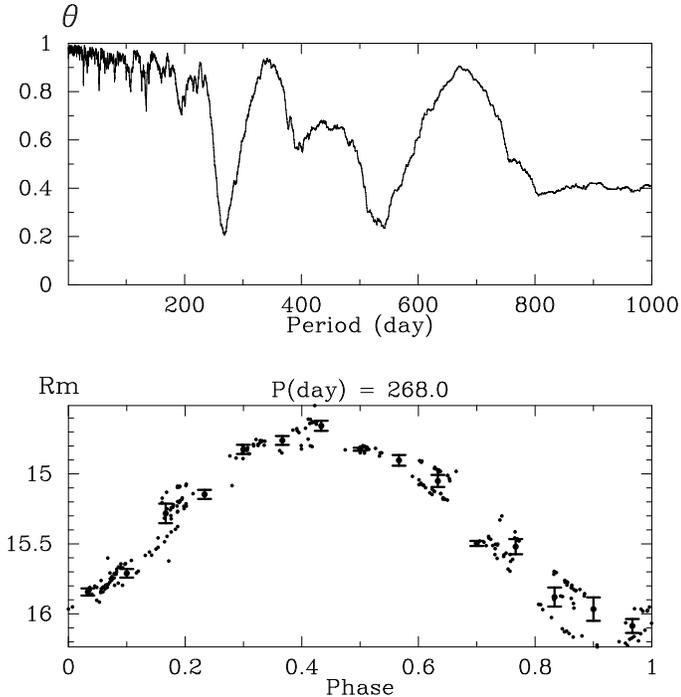,width=9cm}}
   \caption{A sample of PDMM algorithm output for the same star, nlmc1-280227, as shown in Fig.~3. 
	    The top panel shows the reliability of period determination. 
	    The $\theta$ is the sum of the dispersion of data in each bin for 
	    the given period. The period corresponding to the minimum of $\theta$ 
	    is chosen as the most likely period. The bottom panel shows the folded 
	    phase curve with the most likely period. The error bars indicate the
	    standard deviation in each bin. The period of this star was 
	    determined as $P = 268.0$ days with $\theta = 0.21$.}
\label{fig4}
\end{figure}

The total dataset (Series 2 to Series 3) extends over about 3 years
($\sim$1,060 days), so periods less than $\sim$400 days should be well
determined because these have more than 2.5 cycles of variation within
the observation period. We classified 908 candidate light curves by their
features of periodicity.  Among 908 stars, 329 stars show clear
variability and of these, 259 stars had periods which were relatively
well-determined. We rejected light curves like eclipsing binaries,
then chose light curves whose periods were longer than 
$log P \geq 1.5$ (30 days).
We also required $\log P \leq 2.6$
(400 days) reflecting the dataset time span. Finally, 146 light curves
were left. The reliability parameter $\theta$ for these curves is less than 0.67 (the mean $\sim$ 0.38) with 15 folding
bins. We show the distribution of $\theta$ in Fig.~5.

\begin{figure}
  \centerline{\psfig{figure=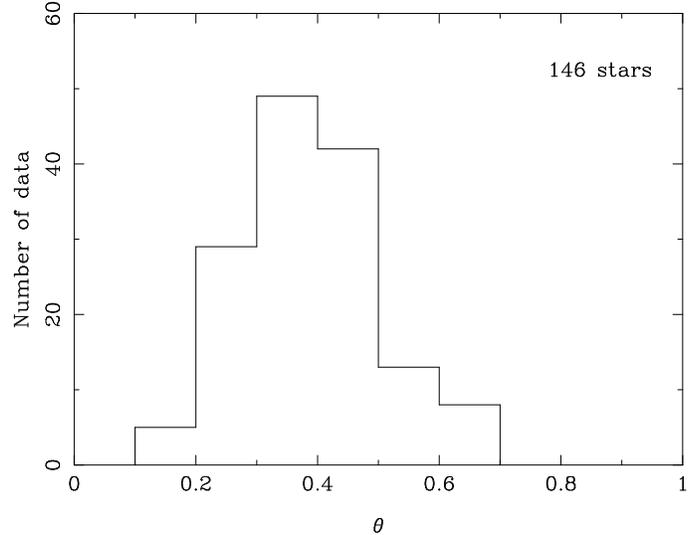,width=9cm}}
  \caption{The distribution of $\theta$ calculated with 15 bins in PDMM. 
	   The mean of $\theta$ is approximately 0.38 for the 146 samples.}
\label{fig5}
\end{figure}

Other candidates, the stars with poor determined period, short period
or small amplitudes, will be the subject of future work.  Here we give
observational values for 146 stars in Table~1 (ID, $\alpha$, $\delta$,
$\langle R_m \rangle$, $\langle V_m \rangle$, $\delta R_m$, $\delta
V_m$, Period, $\theta$). ID is the MOA database star number, ($\alpha,
\delta$) is the right ascension and declination
respectively. Comparing the coordinates of 629 stars in Tables~2
$\sim$ 6 of Hughes \& Wood (1990) with our sample, we identified at
least 34 stars as a control. The ID numbers of their tables are
indicated in the Remarks column.  Spectral types given in their tables are also
shown. The 34 stars consisted of 21 oxygen-rich stars and 13
carbon-rich stars.  In Fig.~6, we show the relation between the
period determined via the PDM method and the period tabulated in Hughes \&
Wood (1990) for 30 stars whose periods are well-determined by us.
It is clear that the periods are coincident and the differences between MOA and Hughes \& Wood (1990) is less than 2 \%.  We used these 30
stars for the conversion to the $K$ magnitude system as described in the following section. Here, we note that Hughes \& Wood (1990) tabulated two
periods, 170 days and 319 days, for the star nlmc3-288533.
We adopted 319 days as their period because we
obtained 307.6 days for the period of this star via our analysis using the PDM method.

\begin{figure}
  \centerline{\psfig{figure=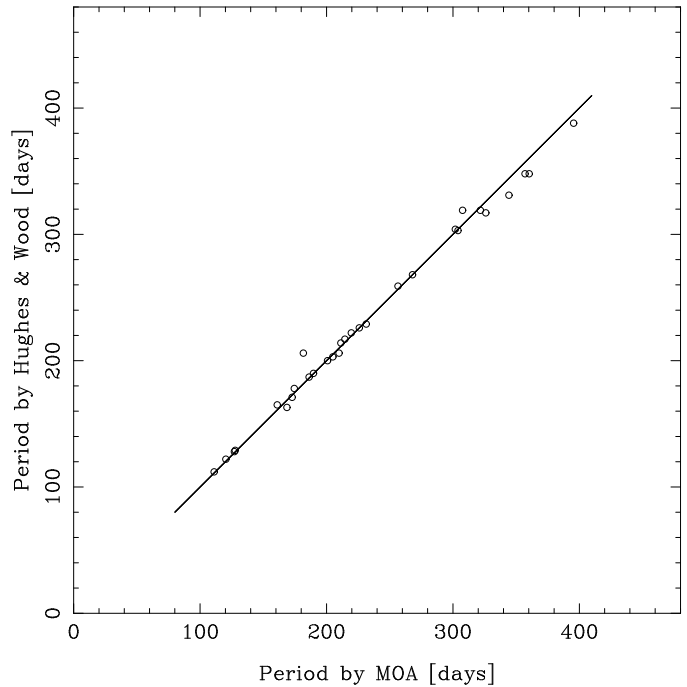,width=9cm}}
  \caption{The comparison between the periods determined by our PDM 
	   (horizontal axis) and the period tabulated in Hughes \& Wood 
	   (1990) (vertical axis). The differences are quite small ($\leq 2 \%$).}
\label{fig6}
\end{figure}

\begin{table*}
   \caption{146 selected long-period red variables}
   \label{catalog}

  \begin{tabular}{@{}lccccccccl}
     MOA ID & $\alpha$ (2000.0) & $\delta$ (2000.0) & $\langle R_m \rangle$ &
     $\langle V_m \rangle$ & $\delta R_m$ & $\delta V_m$ & Period [d] & $\theta$ & Remarks\\

  \hline
nlmc1-108973 & 5:08:35.03 & -69:14:58.03 & 15.43 & 16.34 & 0.85 & 1.09 & 111.3 & 0.38 \\
nlmc1-218899 & 5:08:42.78 & -69:27:37.88 & 15.05 & 16.24 & 2.21 & 3.40 & 225.9 & 0.38 & 0509030-693121 (M) \\
nlmc1-109043 & 5:08:48.65 & -69:07:36.64 & 15.66 & 17.14 & 1.82 & 2.34 & 360.0 & 0.20 \\
nlmc1-324832 & 5:09:21.74 & -69:45:34.53 & 15.81 & 17.77 & 0.68 & 0.94 & 133.5 & 0.54 & (a) \\
nlmc1-49182 & 5:09:21.78 & -69:01:00.58 & 15.12 & 16.26 & 2.40 & 3.62 & 189.7 & 0.23 & 0509382-690440 (M6) \\
nlmc1-49228 & 5:09:37.02 & -69:04:33.08 & 15.35 & 16.71 & 1.67 & 1.97 & 390.1 & 0.47 \\
nlmc1-324861 & 5:10:02.53 & -69:53:24.67 & 15.08 & 16.23 & 0.97 & 1.36 & 203.8 & 0.30 \\
nlmc1-38733 & 5:10:19.99 & -68:58:56.25 & 15.69 & 17.05 & 1.58 & 1.97 & 318.2 & 0.20 \\
nlmc1-153108 & 5:10:32.44 & -69:22:05.18 & 15.56 & 16.58 & 1.12 & 1.49 & 120.3 & 0.43 \\
nlmc1-210097 & 5:10:45.58 & -69:29:18.99 & 15.17 & 16.49 & 2.37 & 3.48 & 214.5 & 0.42 & 0511063-693253 (M) \\
nlmc1-316927 & 5:10:59.42 & -69:52:08.58 & 15.40 & 16.80 & 2.06 & 2.83 & 237.8 & 0.20 \\
nlmc1-99040 & 5:11:01.50 & -69:10:34.26 & 15.06 & 16.14 & 2.23 & 3.46 & 172.8 & 0.26 & 0511195-691408 (M) \\
nlmc1-38993 & 5:11:06.04 & -68:57:43.41 & 16.57 & 17.96 & 0.66 & 0.85 & 136.2 & 0.45 \\
nlmc1-38583 & 5:11:23.97 & -68:58:42.12 & 15.24 & 16.27 & 1.25 & 1.70 & 156.8 & 0.37 \\
nlmc1-316935 & 5:11:26.65 & -69:45:47.39 & 15.43 & 16.74 & 1.02 & 1.24 & 225.6 & 0.41 \\
nlmc1-265052 & 5:11:28.31 & -69:37:48.71 & 15.82 & 16.89 & 1.48 & 2.13 & 233.4 & 0.57 \\
nlmc1-264981 & 5:11:28.64 & -69:35:31.69 & 15.55 & 17.05 & 1.21 & 1.62 & 313.1 & 0.29 \\
nlmc1-38559 & 5:11:42.14 & -68:54:56.73 & 15.71 & 16.90 & 1.83 & 1.74 & 174.5 & 0.37 & 0511580-685826 (M) \\
nlmc1-98891 & 5:11:42.67 & -69:15:01.51 & 15.27 & 16.47 & 2.48 & 3.76 & 186.2 & 0.33 & 0512015-691831 (M) \\
nlmc1-145336 & 5:12:12.00 & -69:23:11.61 & 15.93 & 17.57 & 0.80 & 1.31 & 392.8 & 0.30 \\
nlmc1-200272 & 5:12:17.34 & -69:30:27.11 & 14.97 & 15.93 & 1.50 & 1.96 & 200.8 & 0.25 & 0512385-693356 (M) \\
nlmc1-29759 & 5:12:20.50 & -68:58:37.04 & 15.69 & 16.82 & 1.95 & 2.84 & 120.4 & 0.52 & 0512369-690204 (M) \\
nlmc1-88732 & 5:12:32.79 & -69:07:54.10 & 15.92 & 16.91 & 0.59 & 0.85 & 373.1 & 0.52 \\
nlmc1-29858 & 5:12:54.21 & -68:57:28.94 & 15.10 & 16.35 & 1.18 & 1.44 & 304.4 & 0.27 \\
nlmc1-200302 & 5:13:14.12 & -69:29:30.26 & 15.28 & 16.36 & 1.26 & 1.71 & 155.3 & 0.44 \\
nlmc1-145125 & 5:13:16.04 & -69:16:38.43 & 15.23 & 16.32 & 1.02 & 1.38 & 159.4 & 0.38 \\
nlmc1-145271 & 5:13:21.76 & -69:19:55.66 & 15.66 & 17.31 & 1.08 & 1.50 & 338.8 & 0.30 \\
nlmc1-29756 & 5:13:24.80 & -68:57:00.50 & 15.63 & 17.13 & 1.38 & 1.79 & 329.5 & 0.39 \\
nlmc1-200450 & 5:13:41.55 & -69:29:36.64 & 16.23 & 17.75 & 1.75 & 2.39 & 370.2 & 0.47 \\
nlmc1-29775 & 5:13:48.12 & -69:02:14.16 & 15.43 & 16.68 & 1.71 & 1.95 & 327.9 & 0.62 \\
nlmc1-29830 & 5:13:48.61 & -68:56:03.79 & 15.82 & 17.54 & 1.26 & 1.59 & 375.1 & 0.31 \\
nlmc1-78039 & 5:14:01.66 & -69:06:48.05 & 14.91 & 15.77 & 1.12 & 1.46 & 114.9 & 0.37 \\
nlmc1-189979 & 5:14:16.14 & -69:30:56.46 & 15.47 & 16.87 & 1.46 & 1.93 & 322.2 & 0.32 \\
nlmc1-134678 & 5:14:40.39 & -69:16:23.20 & 15.65 & 17.13 & 1.37 & 1.37 & 331.5 & 0.36 \\
nlmc1-189793 & 5:14:58.35 & -69:34:50.89 & 15.46 & 17.02 & 0.85 & 1.22 & 238.4 & 0.36 \\
nlmc1-19391 & 5:15:11.62 & -68:59:34.56 & 15.57 & 16.88 & 0.91 & 1.11 & 181.7 & 0.36 & 0515286-690250 (M) \\
nlmc1-78061 & 5:15:16.81 & -69:09:55.76 & 15.70 & 17.26 & 1.82 & 1.63 & 360.0 & 0.44 \\
nlmc1-248333 & 5:15:19.16 & -69:44:05.37 & 16.08 & 17.41 & 1.68 & 2.49 & 300.9 & 0.18 \\
nlmc1-78391 & 5:15:26.81 & -69:15:07.48 & 14.56 & 15.77 & 2.67 & 3.53 & 256.5 & 0.27 & 0515461-691822 (M) \\
nlmc1-189766 & 5:15:38.31 & -69:32:27.34 & 15.72 & 17.21 & 1.61 & 2.17 & 203.0 & 0.46 \\
nlmc1-248471 & 5:15:38.71 & -69:40:33.73 & 16.16 & 17.87 & 1.47 & 2.20 & 219.6 & 0.34 & 0516017-694346 (M) \\
nlmc1-68195 & 5:16:06.44 & -69:14:11.00 & 15.15 & 16.25 & 0.95 & 1.26 & 177.4 & 0.53 \\
nlmc1-239264 & 5:16:07.44 & -69:44:25.89 & 15.05 & 16.49 & 1.38 & 2.21 & 129.5 & 0.34 & (b) \\
nlmc1-68289 & 5:16:26.31 & -69:06:26.52 & 16.09 & 17.23 & 1.39 & 1.76 & 106.8 & 0.38 \\
nlmc1-180512 & 5:16:28.90 & -69:34:43.66 & 14.97 & 16.02 & 1.37 & 1.85 & 171.9 & 0.35 \\
nlmc1-10656 & 5:16:41.68 & -68:58:50.43 & 15.42 & 16.57 & 1.10 & 1.46 & 166.5 & 0.40 \\
nlmc1-180542 & 5:16:57.23 & -69:28:40.29 & 15.04 & 16.13 & 1.83 & 2.74 & 169.8 & 0.47 \\
nlmc1-239222 & 5:17:00.56 & -69:37:04.13 & 15.57 & 17.23 & 1.47 & 1.59 & 230.4 & 0.48 \\
nlmc1-180614 & 5:17:25.95 & -69:32:50.35 & 15.29 & 16.33 & 1.71 & 2.07 & 231.4 & 0.35 & 0517478-693554 (C) \\
nlmc1-58105 & 5:17:36.15 & -69:10:28.62 & 15.28 & 16.68 & 2.31 & 2.80 & 254.1 & 0.21 & 0517551-691335 (M) \\
nlmc1-118904 & 5:17:50.20 & -69:19:29.71 & 15.84 & 17.36 & 0.69 & 0.90 & 114.6 & 0.66 & (c) \\
nlmc1-118910 & 5:18:09.10 & -69:19:41.38 & 16.72 & 18.15 & 1.51 & 1.66 & 329.5 & 0.61 \\
nlmc1-228015 & 5:18:12.46 & -69:37:48.46 & 15.70 & 16.83 & 1.50 & 1.89 & 127.8 & 0.46 & 0518353-694048 (M6) \\
nlmc1-280227 & 5:18:14.92 & -69:52:10.99 & 15.29 & 16.47 & 1.56 & 1.86 & 268.0 & 0.21 & 0518400-695513 (C) \\
nlmc1-58139 & 5:19:02.89 & -69:11:29.82 & 15.73 & 17.08 & 1.76 & 1.79 & 323.7 & 0.67 \\
nlmc1-228620 & 5:19:19.20 & -69:37:28.60 & 16.93 & 18.62 & 1.36 & 2.04 & 218.2 & 0.52 \\
nlmc2-189096 & 5:19:19.29 & -69:41:13.12 & 15.23 & 16.44 & 1.72 & 2.62 & 178.4 & 0.33 \\
nlmc1-170484 & 5:19:27.72 & -69:33:31.21 & 15.27 & 16.27 & 1.06 & 1.39 & 127.1 & 0.44 & 0519502-693627 (M6) \\
nlmc2-120429 & 5:19:27.99 & -69:35:12.11 & 15.15 & 16.04 & 1.62 & 2.19 & 200.9 & 0.42 \\
nlmc2-52704 & 5:19:45.24 & -69:15:53.34 & 15.38 & 16.42 & 1.01 & 1.28 & 122.0 & 0.36 \\
\end{tabular}
\end{table*}

\begin{table*}
\contcaption{146 selected long-period red variables}
\begin{tabular}{lccccccccl}

nlmc2-189099 & 5:19:53.23 & -69:41:17.89 & 15.43 & 16.55 & 1.28 & 1.74 & 155.9 & 0.44 \\
nlmc2-177924 & 5:20:19.80 & -69:42:56.70 & 15.55 & 16.90 & 1.73 & 2.24 & 290.1 & 0.31 & 0520437-694548 (C) \\
nlmc2-107253 & 5:20:20.13 & -69:33:44.89 & 15.32 & 16.58 & 1.43 & 1.82 & 326.0 & 0.23 & 0520427-693637 (C) \\
nlmc2-41103 & 5:20:36.32 & -69:23:36.23 & 15.82 & 16.78 & 1.17 & 1.55 & 109.1 & 0.35 \\
nlmc2-242429 & 5:21:05.81 & -69:54:12.77 & 15.71 & 17.49 & 0.69 & 1.17 & 116.9 & 0.45 & (e) \\
nlmc2-177433 & 5:21:21.60 & -69:38:19.96 & 15.77 & 17.28 & 1.46 & 2.21 & 360.3 & 0.40 & 0521450-694107 (C) \\
nlmc2-242076 & 5:21:39.38 & -69:51:48.16 & 15.72 & 17.30 & 0.81 & 1.20 & 100.0 & 0.45 & (d) \\
nlmc2-242294 & 5:21:54.42 & -69:50:23.21 & 15.80 & 16.89 & 1.47 & 2.14 & 110.5 & 0.36 \\
nlmc2-300067 & 5:21:59.83 & -69:55:53.71 & 15.40 & 16.42 & 1.49 & 1.94 & 147.6 & 0.47 \\
nlmc2-166505 & 5:22:26.69 & -69:37:40.10 & 16.36 & 17.46 & 0.98 & 1.13 & 86.3 & 0.44 \\
nlmc2-300177 & 5:22:34.06 & -69:56:21.38 & 15.28 & 16.61 & 1.08 & 1.36 & 304.0 & 0.28 & 0523006-695904 (C) \\
nlmc2-95318 & 5:22:44.35 & -69:33:44.04 & 15.47 & 16.92 & 1.25 & 1.61 & 349.6 & 0.62 & 0523078-693626 (C) \\
nlmc2-29822 & 5:22:47.09 & -69:24:09.76 & 15.47 & 16.90 & 1.31 & 1.76 & 291.2 & 0.30 \\
nlmc2-165804 & 5:22:50.36 & -69:36:04.41 & 15.18 & 16.33 & 1.12 & 1.44 & 259.4 & 0.36 \\
nlmc2-29809 & 5:22:58.41 & -69:22:05.02 & 15.17 & 16.34 & 0.77 & 0.97 & 207.1 & 0.50 \\
nlmc2-300086 & 5:23:01.57 & -69:59:54.20 & 15.14 & 16.47 & 1.70 & 2.31 & 344.3 & 0.18 & 0523288-700234 (C) \\
nlmc2-166417 & 5:23:01.76 & -69:36:33.08 & 15.90 & 17.05 & 2.50 & 3.61 & 113.1 & 0.53 \\
nlmc2-95223 & 5:23:08.01 & -69:26:18.59 & 15.57 & 17.03 & 1.07 & 1.32 & 374.4 & 0.41 \\
nlmc2-95219 & 5:23:22.49 & -69:25:59.77 & 14.83 & 15.93 & 1.32 & 1.64 & 256.4 & 0.39 \\
nlmc2-300087 & 5:23:35.11 & -69:59:59.44 & 15.34 & 16.74 & 1.27 & 1.62 & 321.7 & 0.28 & 0524023-700237 (C) \\
nlmc2-95417 & 5:23:50.06 & -69:26:59.64 & 16.25 & 17.82 & 1.46 & 1.61 & 365.7 & 0.32 \\
nlmc2-289306 & 5:23:54.19 & -70:02:21.75 & 16.06 & 17.63 & 1.26 & 1.56 & 212.9 & 0.30 \\
nlmc2-289270 & 5:24:03.14 & -69:57:01.83 & 15.72 & 17.28 & 0.94 & 1.24 & 349.6 & 0.59 \\
nlmc2-84657 & 5:24:11.45 & -69:31:22.34 & 15.64 & 16.66 & 1.28 & 1.70 & 130.0 & 0.30 \\
nlmc2-151799 & 5:24:33.90 & -69:37:02.27 & 16.17 & 18.24 & 0.92 & 1.50 & 127.6 & 0.45 & (f) \\
nlmc2-151839 & 5:24:41.01 & -69:38:22.69 & 16.07 & 17.64 & 2.13 & 2.19 & 284.4 & 0.35 \\
nlmc2-19377 & 5:25:04.75 & -69:19:37.68 & 16.03 & 17.89 & 0.73 & 0.99 & 121.1 & 0.40 & (g) \\
nlmc2-220283 & 5:25:26.54 & -69:52:41.04 & 15.45 & 16.52 & 1.18 & 1.66 & 136.0 & 0.29 \\
nlmc2-19266 & 5:25:27.46 & -69:23:44.32 & 15.41 & 16.80 & 1.01 & 1.26 & 207.7 & 0.42 \\
nlmc2-19930 & 5:25:32.02 & -69:24:09.81 & 16.44 & 18.09 & 1.15 & 1.81 & 161.0 & 0.28 & 0525543-692639 (M) \\
nlmc2-10220 & 5:25:44.68 & -69:15:18.16 & 15.58 & 17.09 & 1.42 & 1.71 & 324.7 & 0.30 \\
nlmc2-143085 & 5:26:09.35 & -69:38:20.30 & 15.20 & 16.45 & 1.77 & 2.65 & 210.8 & 0.33 \\
nlmc2-279164 & 5:26:40.45 & -69:59:52.40 & 15.43 & 16.67 & 0.91 & 1.21 & 197.4 & 0.26 \\
nlmc2-74407 & 5:26:53.26 & -69:29:16.22 & 15.25 & 16.32 & 1.19 & 1.50 & 145.3 & 0.33 \\
nlmc2-279125 & 5:26:58.78 & -69:58:32.37 & 15.35 & 16.36 & 0.85 & 1.15 & 106.2 & 0.43 \\
nlmc2-210119 & 5:27:07.83 & -69:47:20.33 & 15.38 & 16.77 & 1.73 & 2.38 & 326.7 & 0.41 \\
nlmc2-10255 & 5:27:15.64 & -69:23:51.38 & 15.28 & 16.43 & 0.63 & 0.90 & 145.0 & 0.44 \\
nlmc2-64540 & 5:28:06.72 & -69:32:27.26 & 15.31 & 16.64 & 1.28 & 1.78 & 211.3 & 0.30 & 0528300-693445 (CS) \\
nlmc2-133102 & 5:28:17.73 & -69:42:02.80 & 15.73 & 17.47 & 0.60 & 1.14 & 203.2 & 0.47 \\
nlmc2-64521 & 5:28:33.77 & -69:29:52.95 & 15.85 & 17.07 & 2.07 & 1.78 & 205.0 & 0.27 & 0528568-693208 (M) \\
nlmc3-315892 & 5:28:45.00 & -70:22:53.81 & 15.24 & 16.33 & 0.89 & 1.21 & 168.7 & 0.34 & 0529167-702511 (M1) \\
nlmc3-88398 & 5:28:46.04 & -69:48:34.53 & 16.01 & 17.46 & 1.32 & 1.73 & 244.1 & 0.37 \\
nlmc3-202382 & 5:28:50.33 & -70:00:40.08 & 15.15 & 16.18 & 0.99 & 1.16 & 171.6 & 0.52 \\
nlmc2-266041 & 5:28:51.17 & -70:00:40.45 & 15.10 & 16.16 & 1.00 & 1.14 & 155.4 & 0.54 \\
nlmc3-37184 & 5:28:54.49 & -69:40:12.48 & 15.38 & 16.51 & 1.32 & 2.00 & 157.6 & 0.56 \\
nlmc3-202460 & 5:29:01.01 & -70:06:46.74 & 15.65 & 17.09 & 1.65 & 1.53 & 395.5 & 0.46 & 0529303-700902 (C) \\
nlmc3-262474 & 5:29:09.79 & -70:12:38.77 & 15.28 & 16.52 & 1.21 & 1.43 & 262.3 & 0.51 \\
nlmc3-144789 & 5:29:13.07 & -69:50:42.18 & 15.50 & 17.04 & 1.22 & 1.67 & 355.6 & 0.43 \\
nlmc3-88237 & 5:29:17.03 & -69:43:33.34 & 15.22 & 16.58 & 2.42 & 3.50 & 239.3 & 0.27 \\
nlmc3-37148 & 5:29:22.54 & -69:36:11.89 & 15.89 & 17.49 & 1.58 & 2.30 & 209.9 & 0.29 & 0529467-693825 (MS) \\
nlmc3-37121 & 5:29:29.15 & -69:32:51.15 & 15.34 & 16.46 & 1.92 & 2.95 & 172.6 & 0.30 \\
nlmc3-145171 & 5:29:40.22 & -69:57:37.65 & 16.45 & 18.00 & 2.26 & 2.54 & 316.8 & 0.61 \\
nlmc3-144814 & 5:29:47.07 & -69:53:24.80 & 15.60 & 17.09 & 0.92 & 1.32 & 195.1 & 0.49 \\
nlmc3-88048 & 5:29:52.38 & -69:44:09.13 & 15.33 & 16.29 & 1.27 & 1.62 & 131.2 & 0.20 \\
nlmc3-37227 & 5:29:59.70 & -69:31:41.76 & 15.95 & 17.79 & 0.70 & 1.13 & 264.6 & 0.64 \\
nlmc3-306907 & 5:30:01.02 & -70:20:04.46 & 15.99 & 17.62 & 3.30 & 2.83 & 356.5 & 0.27 & 0530323-702216 (M6) \\
nlmc3-251030 & 5:30:17.43 & -70:18:14.96 & 15.88 & 17.81 & 0.76 & 1.18 & 350.5 & 0.44 \\
nlmc3-306731 & 5:30:26.09 & -70:22:50.30 & 15.99 & 17.07 & 1.41 & 1.60 & 231.1 & 0.22 \\
nlmc3-191203 & 5:31:22.09 & -70:00:30.54 & 16.32 & 17.81 & 1.50 & 1.27 & 384.2 & 0.49 \\
nlmc3-77881 & 5:31:22.87 & -69:46:29.21 & 15.20 & 16.42 & 1.04 & 1.42 & 287.7 & 0.26 \\
\end{tabular}
\end{table*}

\begin{table*}
\contcaption{146 selected long-period red variables}
\begin{tabular}{lccccccccl}

nlmc3-77928 & 5:31:28.35 & -69:40:48.78 & 15.60 & 16.91 & 1.42 & 1.81 & 242.1 & 0.36 \\
nlmc3-306656 & 5:31:33.65 & -70:27:53.96 & 15.77 & 16.82 & 1.70 & 2.51 & 111.1 & 0.40 & 0532066-703000 (M) \\
nlmc3-78091 & 5:31:53.61 & -69:47:07.41 & 15.74 & 17.02 & 1.34 & 1.86 & 162.2 & 0.40 \\
nlmc3-18475 & 5:32:02.11 & -69:34:59.72 & 15.73 & 17.43 & 1.00 & 1.53 & 257.8 & 0.37 \\
nlmc3-241225 & 5:32:02.93 & -70:16:01.83 & 15.74 & 17.41 & 1.10 & 1.48 & 335.4 & 0.31 \\
nlmc3-18489 & 5:32:11.61 & -69:35:45.07 & 15.11 & 16.42 & 2.44 & 3.50 & 246.0 & 0.19 \\
nlmc3-122522 & 5:33:24.88 & -69:59:40.53 & 15.47 & 16.70 & 1.49 & 1.91 & 262.3 & 0.22 \\
nlmc3-179656 & 5:33:46.70 & -70:06:52.06 & 16.01 & 17.40 & 1.37 & 1.52 & 357.5 & 0.48 \\
nlmc3-169613 & 5:34:09.26 & -70:09:14.16 & 15.33 & 16.48 & 0.90 & 1.27 & 190.7 & 0.65 \\
nlmc3-10545 & 5:34:11.85 & -69:38:20.19 & 15.60 & 17.30 & 0.92 & 1.19 & 261.8 & 0.41 \\
nlmc3-288613 & 5:35:04.46 & -70:21:59.56 & 15.95 & 17.60 & 1.29 & 1.56 & 357.1 & 0.32 \\
nlmc3-288533 & 5:35:11.73 & -70:22:43.25 & 15.88 & 17.53 & 1.40 & 1.59 & 307.6 & 0.41 & 0535442-702433 (C) \\
nlmc3-288854 & 5:35:15.43 & -70:22:19.52 & 16.28 & 17.90 & 1.10 & 1.53 & 209.0 & 0.35 \\
nlmc3-222606 & 5:36:03.70 & -70:13:43.72 & 15.49 & 16.94 & 1.43 & 1.79 & 301.9 & 0.29 & 0536347-701529 (C) \\
nlmc3-222499 & 5:36:25.59 & -70:19:15.90 & 15.37 & 16.61 & 1.77 & 2.92 & 173.0 & 0.35 \\
nlmc3-4366 & 5:36:32.90 & -69:31:50.43 & 15.94 & 17.33 & 1.02 & 1.34 & 135.4 & 0.50 \\
nlmc3-162051 & 5:36:44.15 & -70:09:44.26 & 15.01 & 16.74 & 1.64 & 2.22 & 378.2 & 0.41 \\
nlmc3-54070 & 5:36:52.50 & -69:48:14.15 & 15.69 & 17.38 & 0.70 & 1.31 & 235.6 & 0.41 & 0537193-694955 (MS) \\
nlmc3-280668 & 5:37:04.69 & -70:26:27.05 & 15.30 & 16.81 & 1.88 & 2.78 & 238.2 & 0.27 \\
nlmc3-222737 & 5:37:05.26 & -70:18:42.66 & 15.39 & 16.62 & 1.33 & 1.71 & 258.9 & 0.19 \\
nlmc3-272280 & 5:38:23.36 & -70:23:32.55 & 16.22 & 17.68 & 1.48 & 1.96 & 342.2 & 0.18 \\
nlmc3-214306 & 5:38:25.79 & -70:18:21.96 & 15.34 & 16.85 & 0.70 & 1.09 & 241.6 & 0.38 \\
nlmc3-214266 & 5:38:52.12 & -70:18:56.05 & 13.80 & 14.69 & 0.84 & 1.22 & 113.4 & 0.38 & (1) \\
nlmc3-214430 & 5:39:07.31 & -70:17:05.09 & 15.76 & 17.15 & 1.31 & 1.66 & 299.6 & 0.20 \\
nlmc3-214415 & 5:39:16.10 & -70:15:56.36 & 16.10 & 17.74 & 1.21 & 1.56 & 315.4 & 0.23 \\
nlmc3-272161 & 5:39:22.78 & -70:22:27.14 & 15.26 & 16.30 & 1.56 & 2.32 & 129.0 & 0.35 \\
nlmc3-272613 & 5:39:27.49 & -70:25:27.95 & 16.55 & 18.02 & 1.14 & 1.53 & 308.4 & 0.40 \\
  \end{tabular}

  \medskip

Note: $\langle R_m\rangle$, $\langle V_m\rangle$ are the intensity mean magnitude 
of $R_m$ or $V_ m$ respectively, and $\theta$ is the PDM algorithm reliability parameter. (a)$\sim$(g) and (1) are the stars referred to in the text. 
Numbers in the Remarks column are the identification number (with spectral type) in Hughes and Wood (1990).

\end{table*}

\section{Colour Magnitude Diagram (CMD)}

Fig.~2 shows the distribution of our sample of 146 variable stars on a
colour-magnitude diagram (CMD) with 9,050 randomly chosen stars for reference. In this figure, the 146 stars were
located at the tip of the Giant Branch. According to the stellar
evolution theory, stars with initial mass of about $1 - 8 M_{\odot}$
evolve into Asymptotic Giant Branch (AGB) stars with carbon and oxygen
cores. Although we only required a large variation in each light curve
and restricted the colour condition and period in our selection criteria
for variable stars (Section 6), these stars were located in one
particular region in the CMD. Consequently we presumed that these
stars form a group which were in a common evolutionary stage and
pulsate by a common mechanism. However there was an extremely bright
star (indicated by a $\oplus$ mark) in this CMD, nlmc3-214266, which
we have mentioned below. Also we will describe the seven stars with +
marks in Section 11. nlmc3-214266 and these seven stars are denoted by
(1) or (a) $\sim$ (g) respectively in Table~1.

\section{Period-Luminosity Relation}

We show a plot of $\langle V_m \rangle$ magnitude versus period in
Fig.~7.  Open circles indicate stars whose $R_m$ amplitudes 
($\delta R_m$) are smaller than 2.2 mag. Filled circles indicate the
stars whose amplitude are larger than 2.2 mag. The line indicates
the regression fit for the stars with $\log P \geq 2.2$;

\begin{equation}
\langle V_m \rangle = 2.01 (\pm 0.43) * \log P + 12.10 (\pm 1.05)
\end{equation}

\begin{figure}
  \centerline{\psfig{figure=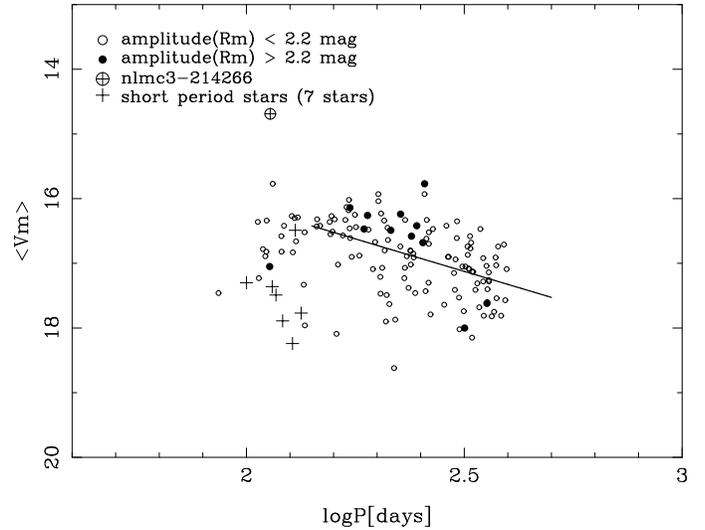,width=9cm}}
   \caption{The Period-Magnitude ($\langle V_m \rangle$) diagram. 
	    Open circles are the stars whose amplitude ($\delta R_m$) are 
	    less than 2.2 mag. Filled circles are the stars whose amplitude 
	    are larger than 2.2 mag. The extremely bright star, nlmc3-214266 
	     is indicated by 
	    the $\oplus$ mark. This star is much brighter than the other stars
            with similar period in this diagram. The cross marks indicate the stars 
	    of group $B$ (indicated in Section 11). The downward line 
	    indicates the least square fitting for the stars with $\log P \geq 2.2$.}
\label{fig7}
\end{figure}

The trend to be fainter in $\langle V_m \rangle$ has been known
in previous studies. The star, which was extremely bright in the CMD,
nlmc3-214266 ($\langle V_m \rangle = 14.69$, $\log P = 2.05$, $\oplus$
mark), was also much brighter than the other stars with similar
periods in this diagram. 
We also show the ($\log P, \langle R_m \rangle$) plot in Fig.~8. 
The slope of this distribution for the stars with $\log P \geq 2.2$
was less than that in ($\langle V_m \rangle, \log P$) plane but still
had a downward tendency. The best fit line for these stars is;

\begin{equation}
\langle R_m\rangle = 1.13 (\pm 0.33) * \log P + 12.84 (\pm 0.80)
\end{equation}

\begin{figure}
  \centerline{\psfig{figure=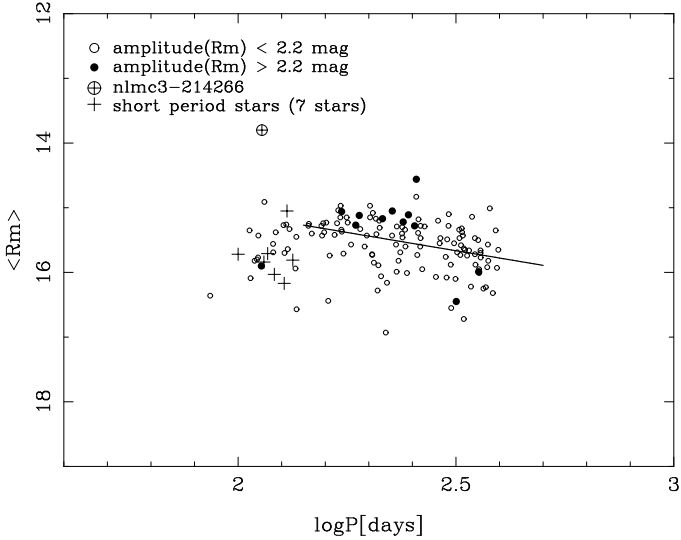,width=9cm}}
  \caption{The Period-Magnitude ($\langle R_m \rangle$) diagram. See Figure 7 for an explanation of the symbols used.}
\label{fig8}
\end{figure}

The star, nlmc3-214266, was again quite bright ($\langle R_m \rangle =
13.80$) and was separate from the other samples. The scatter of
$\langle V_m \rangle$ magnitude ($\sigma \sim 0.52$) was larger than
that of $\langle R_m \rangle$ ($\sigma \sim 0.39$), which was possibly
due to the difference between oxygen-rich stars and carbon-rich
stars. We will discuss this in Section 10.

\section{Period-Colour Relation}

We show the period-colour ($\log P, \langle V_m \rangle - \langle R_m \rangle$)
plot in Fig.~9. The upward line indicates the regression
line for the stars with $\log P \geq 2.2$;
\begin{equation}
\langle V_m\rangle - \langle R_m\rangle = 0.88 (\pm 0.16) * \log P -
0.74 (\pm 0.38)
\end{equation}

\begin{figure}
  \centerline{\psfig{figure=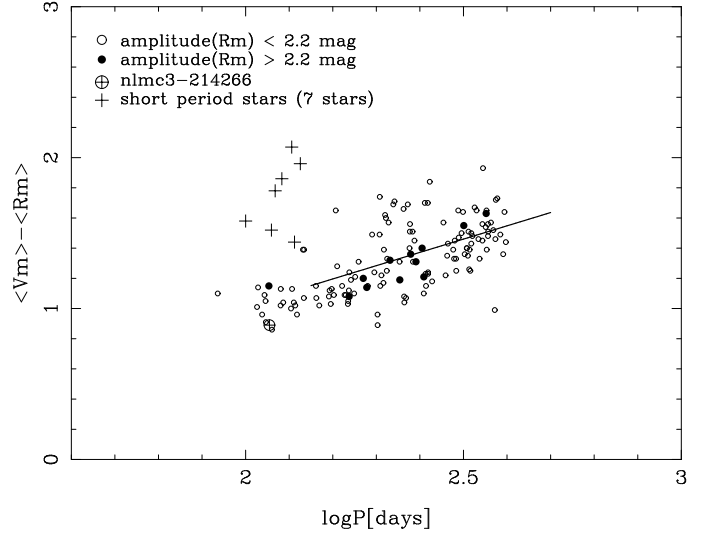,width=9cm}}
   \caption{The Period-Colour ($\langle V_m \rangle - \langle R_m \rangle$) 
	    diagram. 
	    The relationship shows a narrower band compared with Fig.~7 and 8. 
	    The seven stars with short period (+ symbols, see Section 11.2)
	    show the extreme red colour, 
	    while nlmc3-214266 is not conspicuous (see Section 9.1).}
\label{fig9}
\end{figure}

In this figure, the colour of most stars with longer periods ($\log P
\geq 2.2$) were much redder than that of the short period stars, and
the relationship was more constrained ($\delta \sim 0.19$) compared
with that shown in Fig.~7 and Fig.~8.

\subsection{nlmc3-214266}

In Fig.~9, star nlmc3-214266 is positioned in the main
group. Comparing with the other stars of similar period (open 
and filled circles), the effective temperature of this star should be
similar since the colour ($\langle V_m\rangle - \langle R_m\rangle$)
was almost equivalent. The brightness of nlmc3-214266 was high in both
$\langle V_m\rangle$ and $\langle R_m\rangle$, therefore its radius
must have been large compared with the stars in the main group. 
The star was almost certainly HV 1011 which was a suspected Cepheid (Kurochkin 1992). 
Kurochkin was unable to obtain a period for this star and our results do not
conclusively suggest that this star is a classical Cepheid.
When the pulsation period is the same, a larger star must have a higher mass
considering the relation between the period and the mean density of stars.
Therefore we assumed that
nlmc3-214266 was a massive star evolved from the main sequence, and
not an AGB star. The light curve is presented in Fig.~10. The star
nlmc3-214266 was a luminous semi-regular variable.
In the Galaxy, V810 Centauri is one of the variable stars
which is more luminous and has a higher temperature 
than the AGB in the CMD.
Kienzle, Burki, Burnet, and Meynet (1998) estimated that
the spectral type of this star was F8 Ia, and the period was approximately
156 days, $M_{bol} = -8.5$, and the initial mass was $\sim$ 25 $M_{\odot}$. 
Such a heavy star will become a supernova rather than a AGB star (e.g. Becker, 1998).

\begin{figure}
  \centerline{\psfig{figure=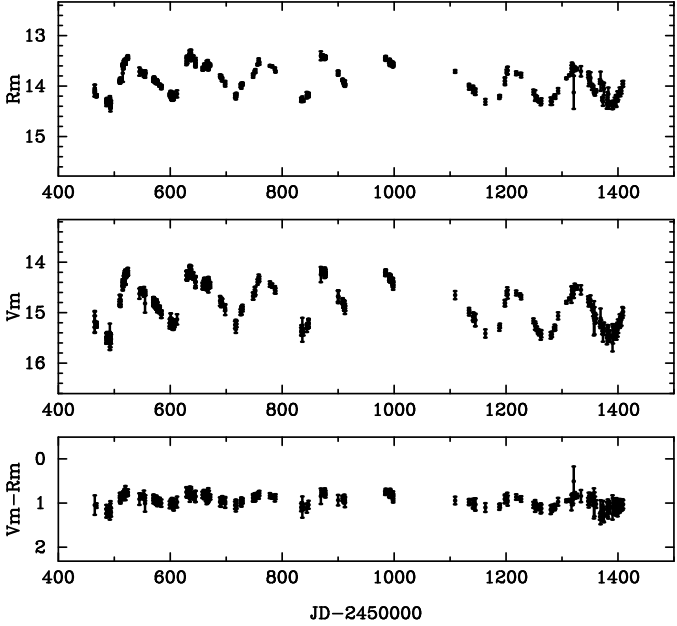,width=9cm}}
   \caption{The light curve of nlmc3-214266. This star is quite bright 
	    in the period-magnitude diagrams (Fig.~7, 8), not red in the 
	    period-colour diagram (Fig.~9), and again bright in the 
	    ($\log P, K_m$) diagram (Fig.~15).}
\label{fig10}
\end{figure}

\subsection{Short Period Red Stars }
    In Fig.~9, several stars whose periods were $\log (P/d) < 2.15$
    were distributed towards the red region and were separated from the
    main group of stars.
    If the colours are different between stars with a similar period and
    luminosity, their mass or pulsation property ($Q = P \sqrt{\bar{\rho}}$)
    should be different.
    It is quite important to investigate their luminosity because
    the relation between luminosity and colour gives information
    about their radius, i.e., mass and pulsation parameter ($Q$) of each star.
    Although our $\langle V_m \rangle$ and $\langle R_m \rangle$ are trustworthy
    they cover too small a wavelength range to enable
    the energy distribution of such red stars to be discussed directly.
    Therefore we study the relationship between our magnitudes and
    the near-IR luminosity in Section 10.

\subsection{nlmc1-239264}
   We note that there were no data points for the minima of the light
   curve for nlmc1-239264. Therefore the mean magnitudes, $\langle V_m
   \rangle$ and $\langle R_m \rangle$, for this star were calculated by using
   only the bright phases. This gives just the brighter limit of the mean magnitude for this star.
   Regarding the period, we assumed that the determined period for the
   star ($P = 129.5$ d, or $\log P = 2.11$) was accurate because there were
   three maxima in the light curve.
   On Figs.~7 and 8, the star was located near the extension of the least
   squares fitted line for stars with $\log P \geq 2.2$.
   When we used a dataset covering complete
   cycles, the stars were located at fainter positions.
   We used $\langle V_m\rangle - \langle R_m\rangle = 1.44$ as the blue or
   lower limit of the colour average in Fig.~9.
   The star was located about half a magnitude redder than the least squares
   fitted line in Fig.~9.
   Considering the incompleteness of our dataset, we may assume that
   the star nlmc1-239264 should be located at a fainter position on Figs.~7
   and 8, and at a redder or higher position on Fig.~9.


\subsection{nlmc3-306907}
    The star, nlmc3-306907, is the remarkable star because of its
    unique colour ($\langle V_m \rangle - \langle R_m \rangle$)
    variation, i.e., the colour was bluer when $\langle V_m\rangle$
    and $\langle R_m\rangle$ were fainter.
    The light variation is shown in Fig.~11.
    Among 146 samples, nlmc2-151839 and nlmc3-191203 also showed the
    same property, though to a lesser extent.
    The colour variations of all the other stars were fainter when the
    $\langle V_m \rangle$ and $\langle R_m \rangle$ were brighter,
    which is a common feature of radial pulsators.
    Such an extraordinary property might indicate the development
    of a particular molecular band caused by a different chemical composition
    in the stellar surface.

\begin{figure}
  \centerline{\psfig{figure=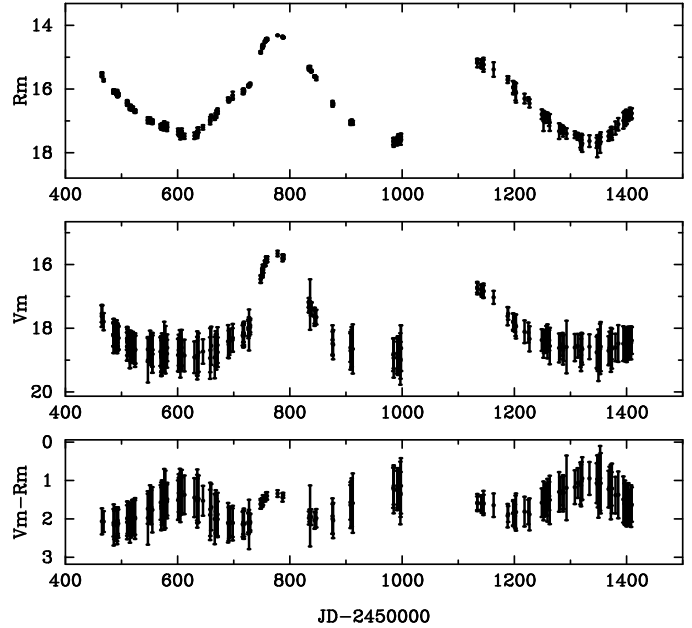,width=9cm}}
  \caption{The light curve of nlmc3-306907.
           The colour ($\langle V_m\rangle - \langle R_m\rangle$) is bluer
           when the $\langle R_m\rangle$ and $\langle V_m\rangle$ are fainter.}
\label{fig11}
\end{figure}

\section{PERIOD TO $K$-MAGNITUDE RELATION}

It is well known that Miras fall on a narrow PL sequence on the
($\log P, K$) plane (Feast et al. 1989; Hughes \& Wood 1990). Wood \&
Sebo (1996) discovered a second sequence located on the short period
side of Miras, while Bedding \& Zijlstra (1998) found strong evidence
that Semi-Regulars (SRs) near the Sun's position in the ($\log P, K$)
plane lie in a region which coincides with the second sequence found
by Wood \& Sebo (1996). Also the existence of at least two (or four)
sequences other than Miras was indicated from the analysis of MACHO
database (Wood 2000).

The $K$-band ($\sim 2.2 \mu$m) corresponds closely to the maximum radiation
flux of long-period red variables, so that the discrepancy between
$K$-magnitudes and bolometric magnitudes should be relatively small. Because
previous studies have been done with $K$ magnitudes, we have to
convert
our observational magnitude $\langle V_m \rangle, \langle R_m \rangle$  into
$K$-magnitudes in order to compare our results with the established PL
relation.

First, to define our conversion scale, we identified 34 stars which were the same
as those published by Hughes \& Wood (1990) from our 146 sampled long-period
red variable stars (see Section 6). The 34 identified stars consisted of 21
oxygen-rich (M-type) stars and 13 carbon-rich (C-type) stars. Furthermore, we
selected 19 oxygen-rich stars whose periods were well-determined in our analysis, and
used them to calibrate our $V_m, R_m$ magnitudes to $K$ values;

\begin{equation}
\langle R_m\rangle - K_{shv} = 2.10 * (\langle V_m\rangle - \langle R_m\rangle) + 1.22
\label{Km}
\end{equation}

$K_{shv}$ was the $K$-magnitude tabulated in Hughes \& Wood (1990)
($shv$ is the prefix of the star number). Hereafter we write $K_m$ in
place of $K_{shv}$ as our custom $K$ magnitude. We discuss here a few reasons
for the scatter ($\sigma \sim0.15$ mag) in this relation. Firstly, and possibly
most importantly, the $K$-magnitudes obtained by Hughes \& Wood (1990) were
determined with only a single observation, whilst we were dealing
with the average magnitudes of at least one hundred measured points. Although
amplitudes of long-period red variable stars are large ($\sim 0.6$ mag) in the
$K$-band, a single epoch observation provides the magnitudes on the various
phase of light variations, so it should have caused extra scattering in our $K_m$
conversion. Secondly, our transformed magnitudes ($V_m$ and $R_m$) which were
obtained with our broad band filters do not correspond to a standard V or
$R_{KC}$ (also $K$) exactly. Finally, as mentioned in Section 4, there was an
uncertainty of approximately $\pm 0.015$ mag dependent on the image position
on each CCD chip.
The result from comparing the $K$ magnitudes of Hughes \& Wood (1990)
and the average MOA magnitudes had a similar scatter to that due to 
the intrinsic variability of the stars.  
We also calculated the probable $K$ magnitudes by extrapolating the
observed light curves backwards to the time of observation of Hughes
\& Wood (1990). The scatter found in this case was larger and most probably
due to the imperfect periodicity of Mira variables.
Since the resulting relation was nearly the same as equation (14) that
expression was used in the following study.

Now we estimated the coincidence of our $K_m$ conversion with formula (14).
In Fig.~12, we show a ($\langle V_m \rangle - \langle R_m \rangle,
\langle R_m \rangle - K_{shv}$) plot for 19 oxygen-rich stars (open circles)
and 11 carbon-rich stars (filled circles) which were common  in Hughes \& Wood
(1990) and our catalogue, and show the well-determined periods by our analysis.
19 oxygen-rich stars  were used to derive formula (14) which is indicated by the
solid line. It is obvious that the conversion formula (14) fits the oxygen-rich
Miras (O-Miras) ($\sigma \sim 0.15$ mag) well but does not fit the carbon-rich
stars ($\sigma \sim 0.34$ mag). The dashed line indicates another conversion
formula using 11 carbon-rich stars:

\begin{figure}
  \centerline{\psfig{figure=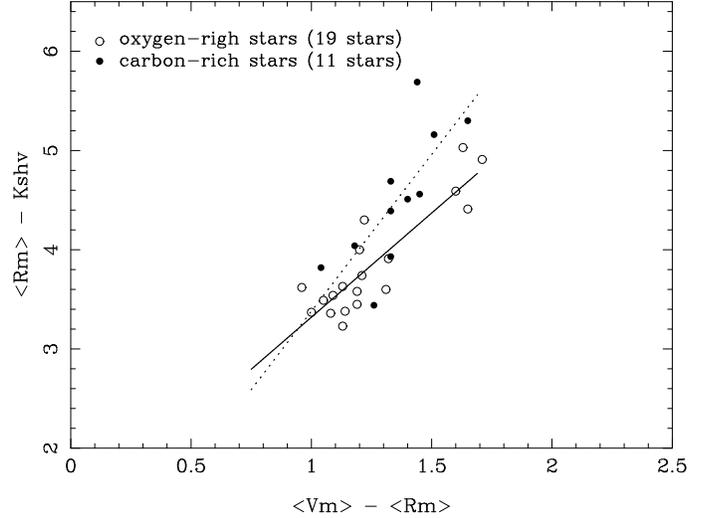,width=9cm}}
   \caption{Comparison of ($\langle R_m \rangle -K_{shv}$) and the MOA 
	    Colour ($\langle V_m \rangle - \langle R_m \rangle$). 
	    $K_{shv}$ is the magnitude as tabulated in Hughes \& Wood (1990). 
	    Open circles indicate the oxygen-rich stars and filled circles 
	    indicate the carbon-rich stars. The solid line indicates the $K_m$ 
	    conversion formula (14), while the dashed line indicates the least 
	    squares fitting for the 11 carbon-rich stars (formula (15)). 
	    The solid line makes a good fit to the oxygen-rich stars, while 
	    the 11 carbon-rich stars fall on the another steeper (dashed) line.}
\label{fig12}
\end{figure}

\begin{equation}
\langle R_m\rangle - K_{shv} = 3.16 * (\langle V_m\rangle - \langle R_m\rangle) + 0.22
\end{equation}

We also show the ($\log P, K_m$) diagram for these 19 oxygen-rich stars and
11 carbon-rich stars in Fig.~13.
The two lines in this figure were the sequences of O-Miras from
previous studies (Feast et al. 1989; Hughes \& Wood 1990). Our
converted magnitudes $K_m$ for 19 oxygen-rich stars agreed well with
these classical relations ($\sigma \sim 0.18$ mag). In contrast, those
for 11 carbon-rich stars appear below the lines ($\sigma \sim 0.25$
mag). According to the results from previous observations, the PL
relation for O-Miras (oxygen-rich Miras) and C-Miras (carbon-rich
Miras) are similar in the $K$-band (Feast et al. 1989). In Fig.~12
and 13, we conclude that our conversion formula (14) makes a good
enough fit to the oxygen-rich stars, but does not fit to the
carbon-rich stars.

\begin{figure}
   \centerline{\psfig{figure=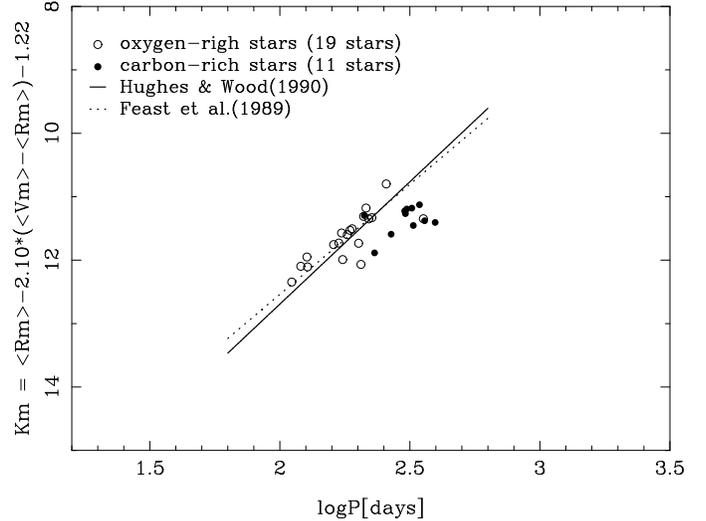,width=9cm}}
    \caption{$K_{shv}$-magnitude versus Period for 19 oxygen-rich stars 
	     and 11 carbon-rich stars which were identified with Hughes \& Wood 
	     (1990) as the same stars. The lines are the O-Mira sequences obtained 
	     by previous studies (Feast et al. 1989; Hughes \& Wood 1990).}
\label{fig13}
\end{figure}

We consider here why such a disagreement arises for the carbon-rich stars. As
is well known, there are roughly two kinds of spectral types, M and C, for
long-period variable stars depending on the relative abundance of oxygen to
carbon in their atmospheres. This causes a remarkable difference in colour
tendency (Smak 1964). The B-band ($V_m$-magnitude in this paper) flux
of oxygen-rich (M-type) stars was smaller than  that for carbon-rich (C-type)
stars. This was due to the existence of some strong TiO bands resulting from
their rich abundance of oxygen. Therefore carbon-rich stars are bluer than
oxygen-rich stars at the same effective temperature, which means that the
value of ($\langle R_m \rangle - K_{shv}$) for carbon-rich stars is greater
than that for oxygen-rich stars even if the colour index ($\langle V_m \rangle
- \langle R_m \rangle$) is the same  (see Fig.~12). This explains why the
$K$-magnitudes for carbon-rich stars were less luminous in Fig.~13, when we
derived the conversion scale, $K_m$, using only oxygen-rich stars.

    The above results were derived from the data by Hughes \& Wood (1990).
    Within these stars (21 O-Miras and 13 C-Miras), 
    the $K$-magnitudes of nlmc3-306907 and nlmc3-288533 
    were reported in Table~4 of Trams et al. (1999).
    The former is an oxygen-rich star and the latter is a carbon-rich star.
    Their results were shown in Table~2 together with $K_{shv}$.

\begin{table*}
  \caption{$K$-magnitudes reported by other observations}
    \label{Other_K}
  \begin{tabular}{@{}lccll}
  Star 		& JD	  & $K$/mag & sources & remarks\\
  nlmc3-306907 	& 2446341 & 10.96   & Hughes and Wood (1990) & phase $\approx$ 0.8\\
		& 2450788 & 10.35   & Trams et al (1999)     & near a light maximum\\
  nlmc3-288533	& 2446341 & 10.58   & Hughes and Wood (1990) & phase $\approx 0.8$\\
		& 2450782 & 10.65   & Trams et al (1999)     & near a light maximum\\
  \end{tabular}
\end{table*}

    Peak-to-peak modulation was found in the light curve of nlmc3-288533.
    The magnitudes tabulated in Table~2 were brighter than 
    $\langle K_m\rangle$, 11.35 and 11.20, of each star.
    The differences of magnitudes at the light maxima
    and $\langle K_m\rangle$ were 1.00 and 0.55 mag respectively.
    Such differences can be accepted as the light variations of this type
    of variable stars.

    When we adopt the $K$-magnitude by Trams et al.(1999), $K_T$, in Fig.~12,
    the position of nlmc3-306907
     moves upward and seems to belong to the carbon-rich stars.
    However, this star should not be regarded as a typical oxygen-rich star
    because of the unusual colour variation it showed as mentioned in Section 9.4.
    The position of nlmc3-288533 in Fig.~12 will
    be slightly lower than the current position.

    Figure 14 shows a similar diagram using the $K$-magnitude
    ($K_S$ : $\sim 2.15 \mu$m) of the DENIS catalogue
    (http://cdsweb.u-strasbg.fr/denis.html).
    Among 146 samples, 139 stars have found, and $K$-magnitudes were given
    for 119 stars in the DENIS catalogue.
    The correlation between $K_S$ and $K_m$ is given by the equation,

    \begin{equation}
    K_S = 0.9 (\pm 0.09) * K_m + 0.73 (\pm 1.07)
    \end{equation}
    with a standard deviation of 0.47.
    Such a scatter may be explained as the results of intrinsic light
    variability.
    In Fig.~14, oxygen-rich and carbon-rich stars were located in the
    lower and the upper regions as seen in Fig.~12.
    The star nlmc3-306907 was again located in the carbon-rich star region.
    No $K_S$ data for nlmc3-288533 has been found.
    If nlmc3-306907 is a typical oxygen-rich star, this indicates that some oxygen-rich
    stars may exist at the upper side of the oxygen-rich relation in Fig.~12.
    Therefore brighter oxygen-rich stars than given by equation (14) may exist.

\begin{figure}
  \centerline{\psfig{figure=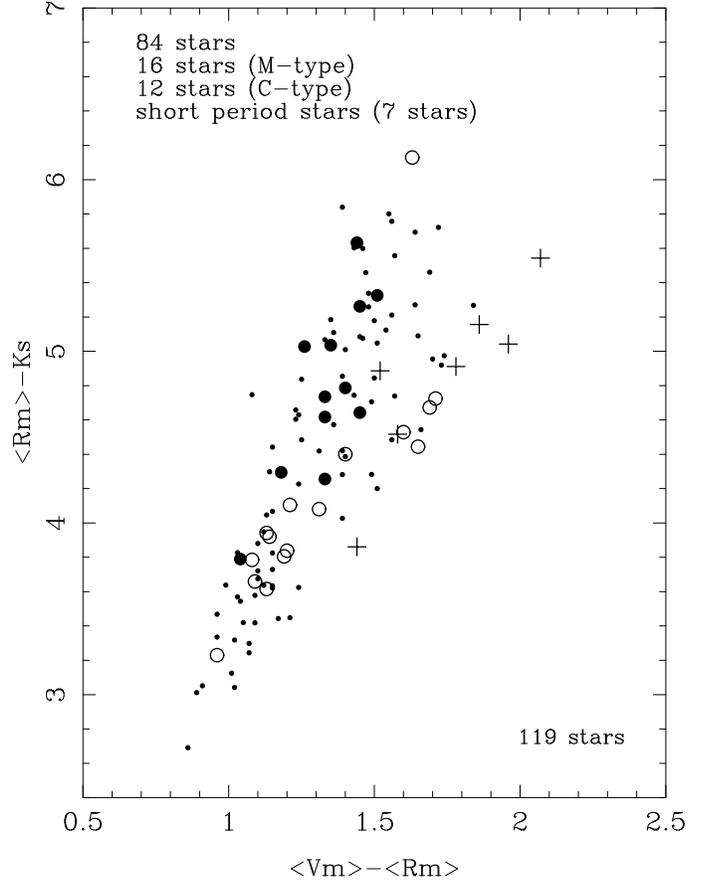,width=9cm}}
  \caption{The ($\langle V_m\rangle - \langle R_m \rangle, \langle R_m\rangle - K_S$)
	   relation for identified 119 stars. $K_S$ indicates the $K$-magnitude
     	   from the DENIS catalogue (see text).}
\label{fig14}
\end{figure}

We show a ($\log P, K_m$) diagram for our sample of 146 stars in Fig.~15. In this figure,
the solid diagonal line is the LMC PL
relation for O-Miras of Hughes \& Wood (1990) and the dotted line shows
the track for O-Miras of Feast et al.  (1989). 
Three groups, designated $A$, $B$ and $C$ are evident in Fig.~15.
They are discussed in detail in Section 11.
The stars in the most populous group lie close to these lines. 
    nlmc3-306907 is located in the same region as carbon-rich stars.
    Another star, nlmc2-64521, is also located below the
    established relation. Therefore it is clearly hard to decide the all
    star below these lines are carbon-rich stars.


One of the 13 carbon-rich stars, nlmc2-64540, lies on these lines as an exception, and the remaining 12 carbon-rich stars fall below these lines. 
The slope of $K_{shv}$-magnitude versus MOA Colour 
(see Fig.~12 and formula (15)) for carbon-rich
stars is steeper than that for oxygen-rich stars, so these lines intersect.  This was why this exceptional
carbon-rich star was located near the line of formula (14) and lies on
the sequences for O-Miras in Fig.~15 (also in Fig.~13). 

We note it was possible that at least 50 $\sim$ 60 carbon-rich stars were
included in the sample of 146 stars
because the ratio of oxygen-rich (the sample of 21 stars) and carbon-rich 
(the sample of 13 stars) stars was approximately 
$\sim 62 \%$ and $\sim 38 \%$ respectively.
These could generate the extra scatter in the ($\log P, K_m$) plot.

\begin{figure*}
  \centerline{\psfig{figure=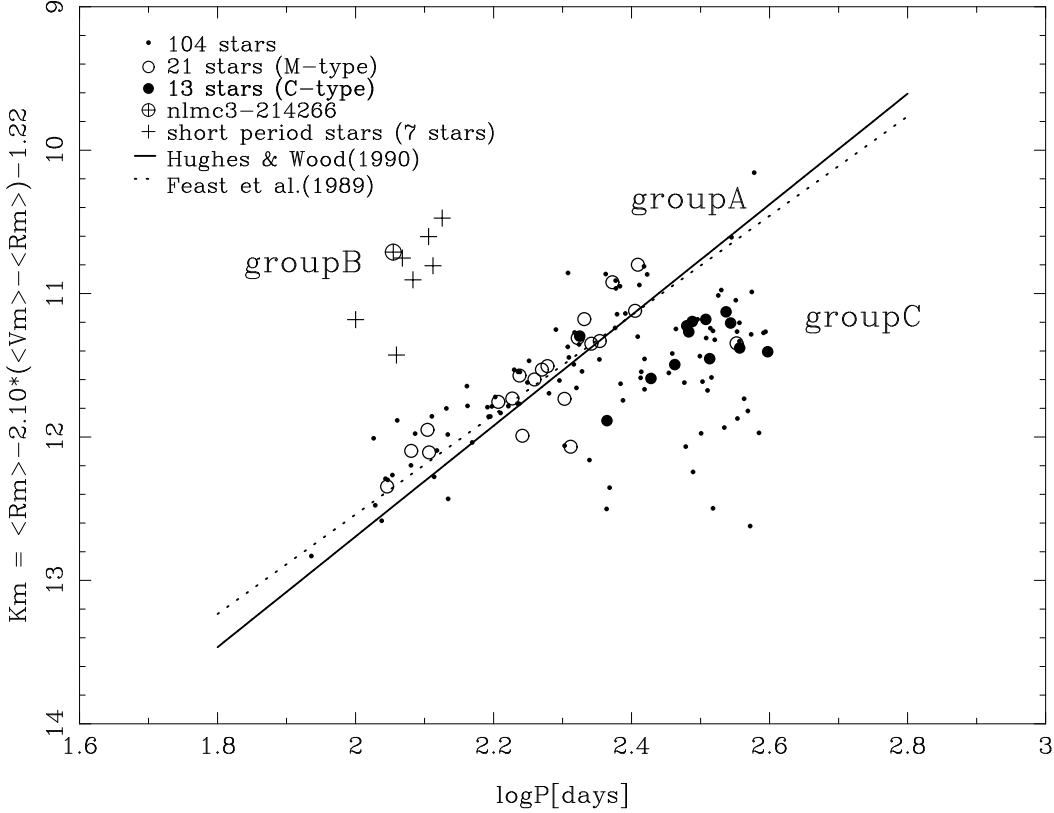,width=14cm}}
   \caption{The ($\log P, K_m$) diagram for 146 long-period red variable stars. 
	    Two lines are the ($\log P, K$) relations for O-Mira sequences by 
	    the previous studies.  The stars in the most populous 
	    group of our samples lie on the O-Mira sequence. One exceptional 
	    carbon-rich star also lies on the O-Mira sequence, while most of 
	    carbon-rich stars fall below these sequences. The + marks 
	    (seven samples) indicate the noteworthy stars with short period, 
	    and the $\oplus$ indicates nlmc3-214266 (see text). An explanation 
	    of groups $A$,  $B$ and $C$ is given in the text.}
\label{fig15}
\end{figure*}

When we converted the MOA data into 
$K$ magnitudes by using equation (15),
most of stars including 11 carbon-rich stars (around group $C$
in Fig.~15) were found near the empirical LMC PL relation for C-Miras,
but some of them remained in the fainter domain of the diagram.
The isolation of the short period variable stars
(seven + mark stars and a $\oplus$ star, defined in the next section)
became clearer in this case.

\section{DISCUSSION}

\subsection{$\log P - K_m$ diagram}

It is difficult to estimate $K$-magnitudes from the large
photometry database as it was originally designed to detect
microlensing events. However we have now succeeded in estimating
$K$-magnitudes using the properties of oxygen-rich and carbon-rich
stars as described in the previous Section. 
Based on these results, we can investigate the ($\log P, K_m$)
diagram more carefully.

As described in Section 10, our conversion of $\langle V_m \rangle$
and $\langle R_m \rangle$ into $K_m$-magnitude could not be applied for
carbon-rich stars and they are located below the main sequence of
Figures 13 and 15. Therefore the scatter in the ($\log P, K_m$) plane
(Fig.~15) must be associated with these stars. 
However, even taking the above fact into account, the scatter in
the plot is too large to be described by one sequence. To demonstrate
this point, in Fig.~16 we show a histogram of vertical distance from
the line of O-Mira sequence by Hughes \& Wood (1990):

\begin{equation}
K = -3.86 * (\log P - 2.4) + 11.5
\end{equation}
The error bars are equal to the square root of the number of stars in each
of the three groups indentified in Figure 16.

\begin{figure*}
  \centerline{\psfig{figure=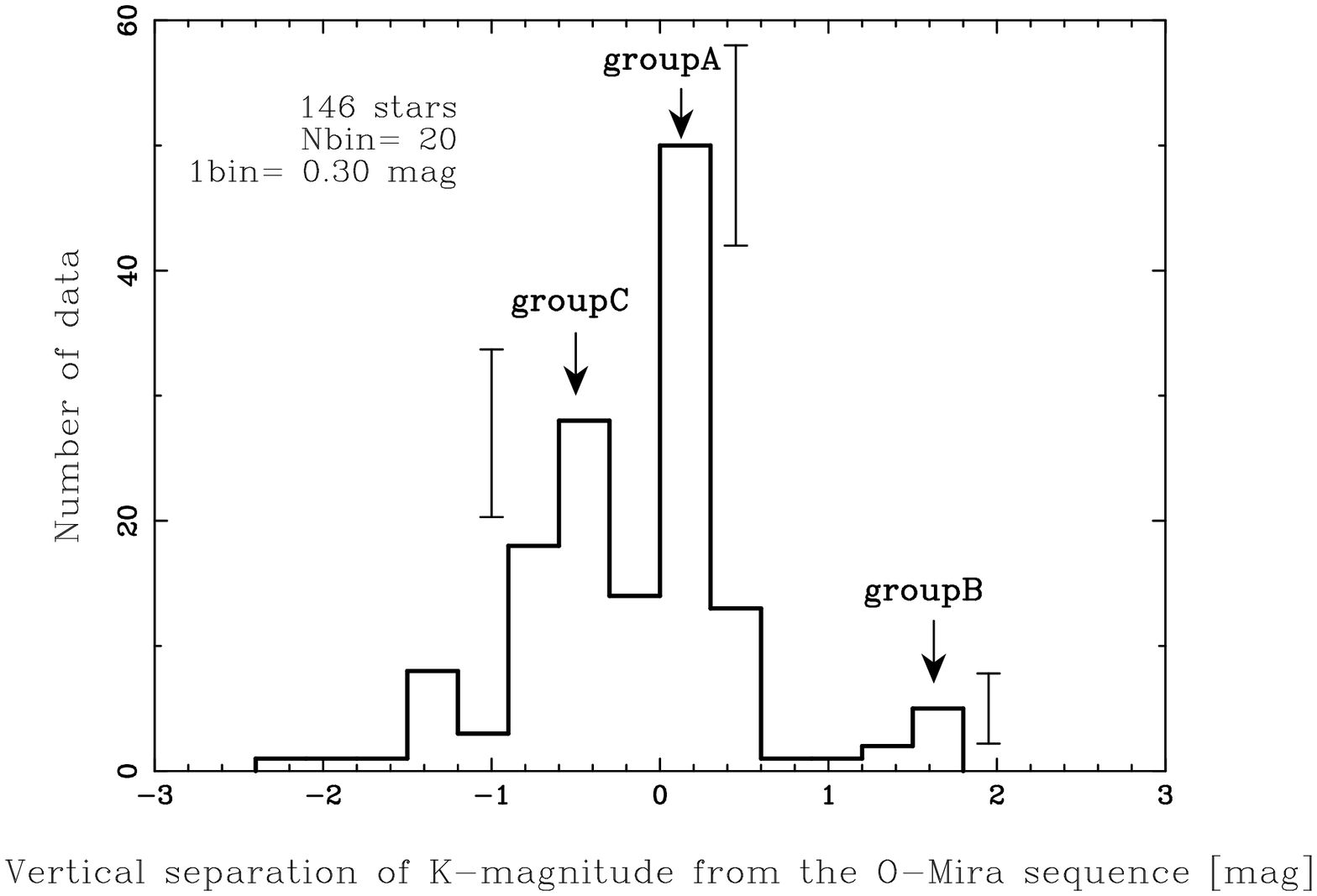,width=9.5cm}}
   \caption{The histogram of the vertical distance of $K$-magnitude 
	    from the O-Mira sequence of Hughes \& Wood (1990) in Fig.15. 
	    The 146 star sample is used. The highest peak at $\delta K = 0$ 
	    contains stars agreeing with the Mira ($\log P, K$) relation by 
	    Hughes \& Wood (1990). We classified these stars as group $A$. 
	    The stars separated by $\sim +1.5$ mag from group $A$ were defined as group $B$. 
	    The stars separated by $\sim -0.6$ mag were defined as group $C$ (see text). 
	    The error bars beside each group are described in the text.}
\label{fig16}
\end{figure*}

A vertical separation of zero means it was on the Hughes \& Wood (1990) track in Fig.~15. The most populous group ,$A$, lies at this position.  This group contained the
stars with largest amplitude. There was a small peak (group $B$, eight
stars) around $+1.5$ mag (corresponding to a leftward shift in
period in Fig.~15). The group $B$ stars are relatively small
amplitude variables. Another peak which is relatively broad can be identified at around $-0.6$ mag (group $C$, corresponding to a rightward shift in
period in Fig.~15). We define each group in Fig.~16 using the 
following limits; the stars distributed between 0 and 0.6 mag as group $A$,
the stars distributed greater than 0.9 mag are as group $B$, and
stars distributed between $-0.9$ and $-0.3$ mag as group $C$.

In Fig.~17 we show the amplitude distribution, $\delta R_m$ for the
stars in each group. In Fig.~18 we show the ($\log P, \delta R_m$) diagram
with the same symbols as Fig.~15.
In Fig.~17, it is clear that most of largest amplitude
($\delta R_m \geq 2.2$ mag) variable stars are  group $A$
stars, which
provides strong proof that these stars are Miras.

\begin{figure*}
  \centerline{\psfig{figure=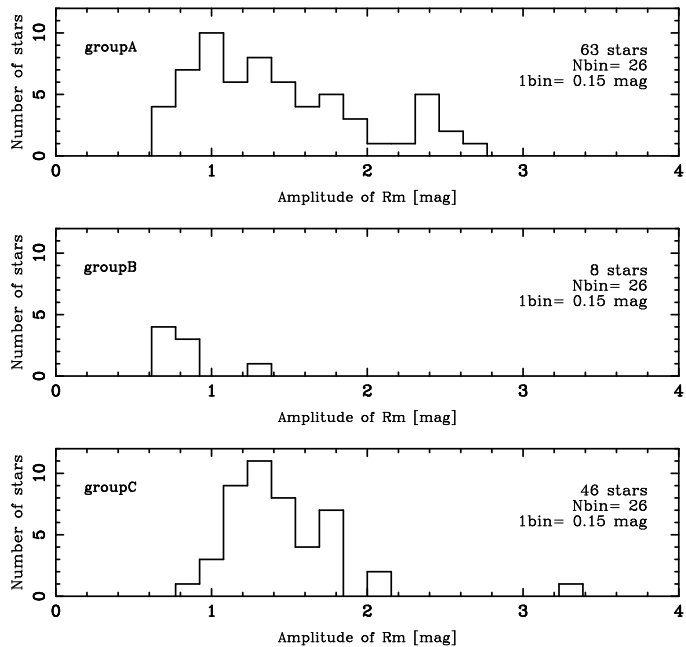,width=9cm}}
   \caption{The distributions of pulsation amplitude in $R_m$ for 
	    each group. The upper panel shows stars on the most populous 
	    sequence (i.e., group $A$). The middle panel shows the distribution 
	    for the stars in group $B$ which were located on the shorter period 
	    side of group $A$. The bottom panel shows that for the stars in the
	    group $C$ 
	    sequence which were located on the longer period side of group $A$.}
\label{fig17}
\end{figure*}

\begin{figure*}
  \centerline{\psfig{figure=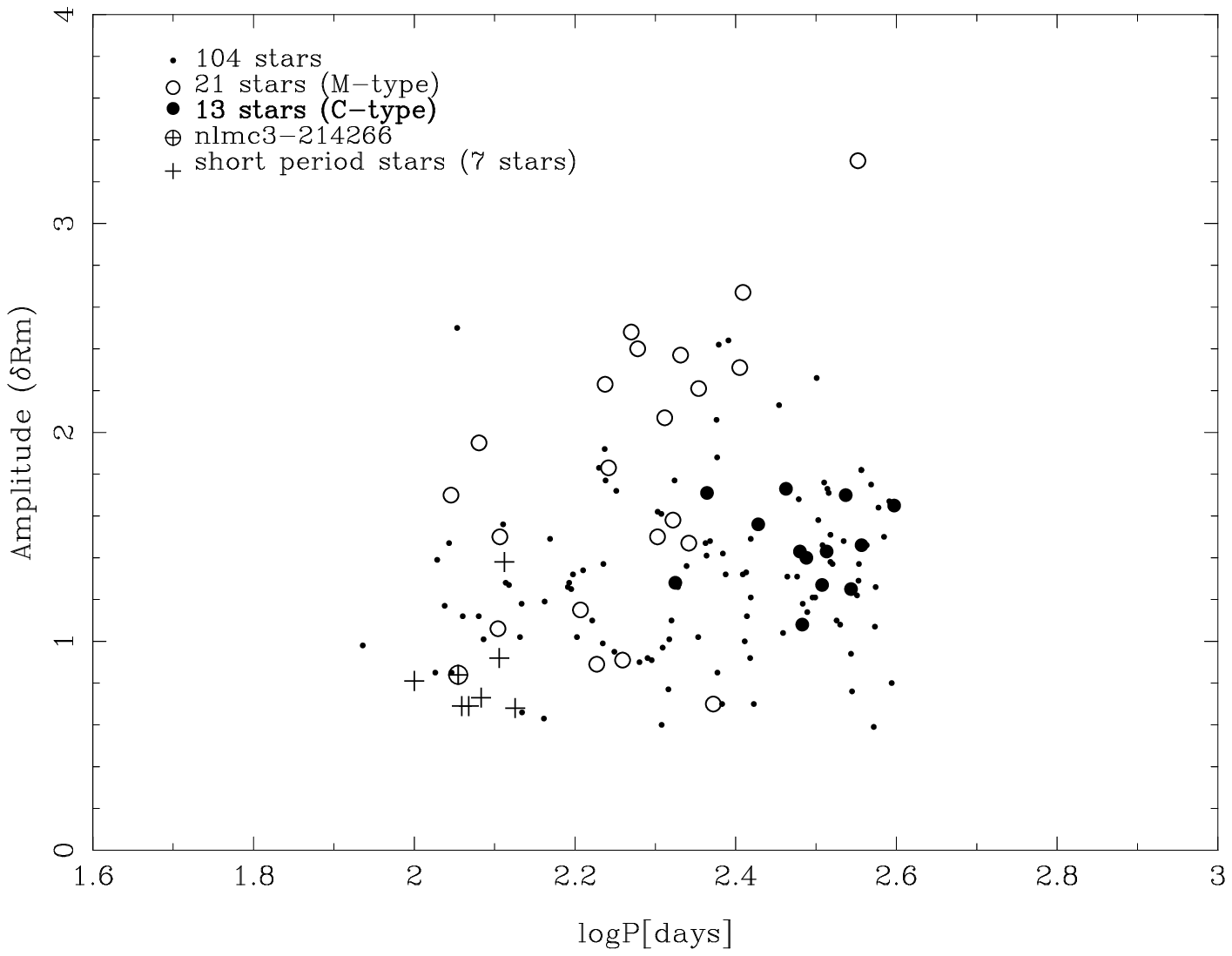,width=12cm}}
   \caption{The period-amplitude ($\delta R_m$) diagram.
	    The symbols are the same as Fig.~15. 
	    Group $B$ stars have been found in the low amplitude domain.}
\label{fig18}
\end{figure*}

In contrast, the amplitudes of stars in group $B$ were smaller
($\delta R_m \leq 1.4$ mag) than most of the others as shown
in Fig.~17 and Fig.~18. This group includes the seven short period stars and nlmc3-214266. This group is located on the very short period side
($\log P \leq 2.1$) of the ($K_m, \log P$) plane (Fig.~15). The group of seven short period stars is located in similar positions in the
period-colour or period-magnitude diagrams (Fig.~7, 8, 9).  
All of these stars showed relatively inferior periodicity: the mean
value of $\theta$ was $\sim 0.43$, compared with the mean of $\theta
\sim 0.38$ for the stars in group $A$.
We obtained periodic variable stars via our selection criteria (Section
5), but we did not strictly eliminate stars whose periodicities were
relatively poor. So the stars in group $B$ were somewhat semi-periodic
compared to the stars in group $A$ (see Fig.~5).


\subsection{The Group $B$ Stars}

   To examine $K$ magnitudes of the short period variable stars,
   we compared our estimates with $JHK$ observations made by the 1.9 m
   telescope at the South African Astronomical Observatory.
   Star identification, magnitudes and date of observations are listed
   in the first five columns of Table~3.
   Mean $K$ magnitudes derived by equation (14) are tabulated in the 6th column.
   The first three stars of the table are stars (a), (b), and (g) of group $B$,
   and the last one was the probably non-AGB star ($\oplus$ in Fig.~15)
   of group $B$ (see Table~1).
   The differences in $K$ mag obtained at the SAAO and derived averaged
   $K$-magnitude, $\langle K_m\rangle$, were 0.33, greater than 1.02, 0.06,
   and 0.04, respectively. Except for the second star, nlmc1-239264, the differences
   do not seem remarkable. We predicted successive light maxima of these
   four stars using the observed light curves, and have found that the
   SAAO observations were done near a predicted light maximum for the
   star nlmc1-324832, and near their predicted light minima for the other
   three stars.
   If their light variations were sufficiently regular, the SAAO 
   $K$-magnitudes of nlmc1-324832 had to be brighter than 10.47.
   This contradicts the observed brightness, so that the regularity
   of the light variation should be suspected.
   For the other three stars, $K$-magnitudes were fainter than
   their $\langle K_m\rangle$.
   The observed magnitude for nlmc1-239264 was fainter than
   our adopted mean magnitude, and the difference was 1.02 mag.
   Such a difference looked too large when compared with the other two stars.
   However, the true difference between $K_{SAAO}$ and the mean $K$ would be
   less than 1.02 mag because the adopted mean magnitude should be brighter
   than true mean magnitude.  
   Therefore the difference may be acceptable with the true amplitude in $R_m$
   greater than 1.38 mag.
   The question remains as to whether these differences originate from
   the uncertainty in $\langle K_m\rangle$ or the semi-regular nature
   of these variables.
   As indicated in Fig.~14, the $K_S$ magnitudes for seven stars in group $B$ are were located along the
   line of the O-Miras.
   This tends to support the use of equation (14) in estimating
   the $K$-magnitudes. Further analysis of our database will resolve
   this question.

\begin{table*}
  \caption{JHK magnitude by SAAO}
    \label{SAAO_JHK}
  \begin{tabular}{@{}lcccccc}
  Star & $J$ & $H$ & $K$ & JD-2450000 & $\langle K_m\rangle$ & $\delta{R_m}$\\
  nlmc1-324832 & 12.18 & 11.12 & 10.80 & 2037.3 & 10.47 & 0.68\\
  nlmc1-239264 & -     & 12.25 & 11.83 & 2037.3 & 10.81 & 1.38\\
  nlmc2-19377  & 12.43 & 11.36 & 10.96 & 2037.3 & 10.90 & 0.73\\
  nlmc3-214266 & 11.74 & 10.92 & 10.75 & 2037.3 & 10.71 & 0.71\\
  \end{tabular}
  \medskip
  
  Note: $J,H,K$ magnitudes are measured to within $\pm$ 0.03 mag.

\end{table*}

    Amongst the eight stars of group $B$, nlmc3-214266 should be a massive
    star, not a star evolved from the red giant branch or the first
    red giant branch as pointed out in Section 9.1.
    We examine the properties of the other seven stars.
    The differences between the brightness of the group $B$ stars and
    of the group$A$ stars is evident in Fig.~15.
    The faintest star, nlmc1-118904, was equally bright in $K_S$ with the
    other six stars of group $B$.
    The amplitude of nlmc1-239264 was larger than the
    other six stars, but the position on the period-luminosity diagram
    was not peculiar. We have to study the nature of this star by
    comparison with more abundant samples to determine whether or not
    this star is different from the others.

In Fig.~15, the stars in group $B$ were located parallel to the sequence 
of group $A$ (corresponding to the Mira sequence of Hughes \& Wood (1990)) 
and separated by $\delta \log P \sim 0.4$. 
    The seven stars in group $B$ were located at a similar position
    to the short-period red (SP-red) stars which were reported by Feast \&
    Whitelock (2000) analyzing Galactic Miras. While the SP-red stars
    had large amplitude and Mira-like properties, our group $B$
    stars had small amplitudes except nlmc1-239264 and showed less regular
    variations, different from the Galactic Miras.
    The definition of the Galactic Miras is that their amplitude in
    standard $V$-light should be larger than 2.5 mag. 
    It is a natural classification because there are few stars
    with such amplitudes.
    However, such a gap was not distinct in $I$-band observations of long
    period variables in the LMC (Whitelock, 1997).
    If we try to define Miras in the LMC by Fig.~17, the boundary might be
    at around 2.2 mag, though the small sample size makes that inconclusive.
    Therefore  we discuss the long period variables of the LMC
    with no sharp distinction between Miras and semi-regulars.

The theoretical models of long-period variables predict
the ratio of fundamental to first or second overtone period should be
$\delta \log P \sim 0.3-0.4$. 
The stars in group $B$ were very similar to those compared with
the higher mode pulsators by Wood and Sebo (1996).
Higher mode pulsators are also expected
to have smaller amplitudes than fundamental mode pulsators.  
The smallness of the amplitudes of these stars
in Fig.~17 and Fig.~18 
supported this point. Therefore if the stars in group $A$ were fundamental
pulsators, the seven stars in group $B$ could be first or
second overtone pulsators.
On the other hand, as described above, 
these stars might be the counterparts of short-period red
(SP-red) stars reported by Whitelock, Marang, and Feast (2000).
These authors point out that the kinematics of such stars were different from
the blue stars with similar periods.
Such differences indicate a considerably different stellar age and
imply a difference of evolutionary process due to a
distinct chemical composition.
A further development in the theory of stellar structure is needed
before it will be possible to judge whether these seven stars are
counterparts of SP-red stars (and are higher mode pulsators) or not.

To study these stars further we plotted, in Fig.~19, the
$\langle V_m\rangle - \langle R_m\rangle$ colour against
$K_m$ magnitude with the symbols being the same as in Fig.~15.
The seven stars of group $B$ are located at the reddest and highest
luminosity region where longer-period variables are also located.
This suggests that the radius of these stars is very similar
to that of longer-period stars.
The difference in the period of  stars of similar radius is explained by the difference in the mass $M$
and the pulsation property $Q$.
The estimate of the pulsation property for these stars has not been
completed because of the uncertainty in the pulsation theory, but
the effect of the chemical composition and the stellar mass
on $Q$ looks negligible for a given mode (Barthes and Tuchman 1994).
The difference in the period-temperature relation found in the
globular clusters (Whitelock  1986) has been resolved using
different chemical compositions, but the difference
in the period of the stars of which the colour is the same
cannot be explained this way.
The difference in the stellar mass and the mode of pulsation should
be studied to explain the difference in the period.
When the distribution of periods with stellar physical properties
is continuous, 
it is reasonable to suppose that this is due to changes in stellar
mass or other stellar parameters such as chemical composition.
Since the periods of these seven stars are separated from
their long-period counterparts, the enhancement of a different
mode is suggested.
Further progress in data reduction is required to examine the
existence of the gap in the periods of this luminous red domain
of the colour-$K$ diagram.

When we consider the excitation of different modes,
we have to study the reason for such a mode selection.
In the classical Cepheids and RR Lyrae stars, the higher mode
pulsators have been found on the blue side, or small stellar radii side,
of the colour-magnitude diagram.
Therefore it seems difficult to compare directly the group $B$ stars
with higher mode pulsators of the Cepheid Instability Strip.
On the other hand, 
it may be interesting to compare the surface gravity between them.
The surface gravity of the higher mode pulsators is high for stars 
in the Cepheid Instability Strip.
This implies that the surface gravity of the group $B$ stars might be also high.
The critical frequency of the stellar atmosphere
(i.e., the boundary frequency of reflection or transmission at the
stellar surface) must be high for a high surface gravity.
The enhancement of high frequency, or short period pulsation in group $B$ 
stars suggests high stellar mass
when we suppose the same situation for the transmission and
reflection for all pulsating stars.

\begin{figure*}
  \centerline{\psfig{figure=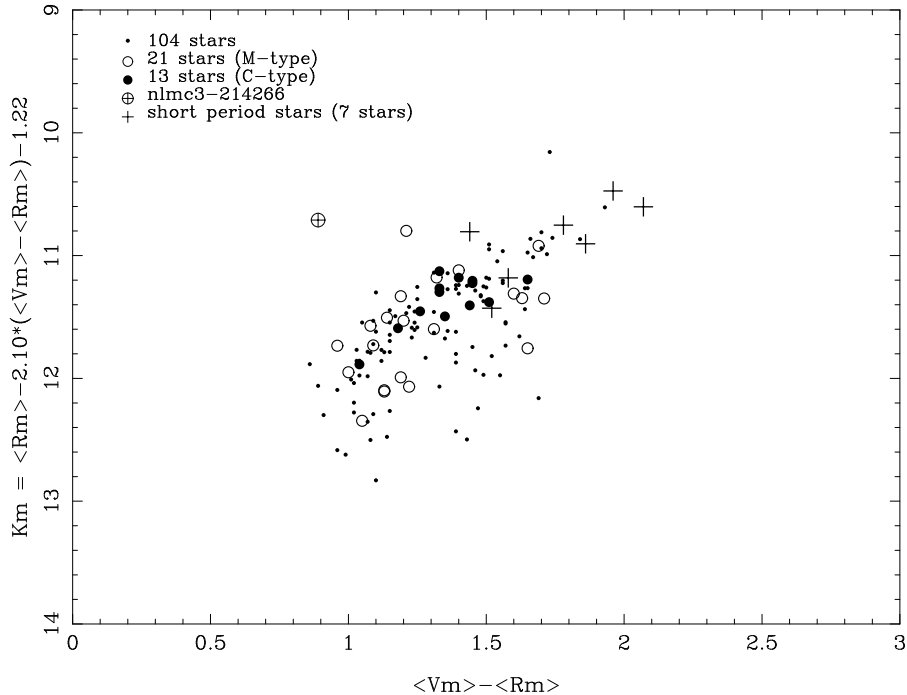,width=12cm}}
   \caption{The ($\langle V_m \rangle - \langle R_m \rangle, K_m$) 
	    diagram for 146 samples. The symbols are the same as Fig.~15. 
	    The seven stars with short period (+) are located at the reddest 
	    and highest luminosity region.}
\label{fig19}
\end{figure*}

\subsection{The Group $C$ Stars}
There are 46 stars in group $C$. Many carbon-rich stars might be included
in this group. 12 group $C$ stars are the same as the carbon-rich stars
tabulated in Hughes \& Wood (1990). There was a possibility that about
50 $\sim$ 60 stars among the sample of 146 are carbon-rich stars as
mentioned in Section 10. These carbon-rich stars were estimated to be
too faint in our $K_m$ conversion. Also, in the amplitude distribution 
(Fig.~17), there was no obvious difference between group $A$ and group $C$.
We conclude most group $C$ stars were carbon-rich and should be included in
group $A$.
However, we need to
distinguish oxygen-rich stars from carbon-rich stars in order to
investigate whether a long-period sequence exists, and if the strip
structure is real or not. Future study of the relationship using
a $B$-band filter should help resolve this issue.

\subsection{Strip Structure}
Wood (2000) presented the sequential structure of the PL diagram using
a large sample. Because our sample size is small, it was hard to
confirm this structure, except for the existence of a separate group 
of red short-period stars.  However, it should be noted that these red
short-period stars were of small amplitude ($\delta R_m$) and less
regular (small $\theta$). These will be useful characteristics in the
study of the nature of these stars.
In the period-$K$ diagram, the stars, nlmc3-214266 and nlmc1-239264, were
also contained in group $B$ implying that it is necessary to investigate
the properties of individual star even in large scale photometry
in order to understand the period-luminosity relation.
The individual star discovered by Wood (2000) should also be checked
to determine whether it is to be included in group $B$ and the in the set of Galactic red SP stars.

\section{SUMMARY}

We selected long-period red LMC variables from the three year MOA
database, and executed a careful period analysis using the PDMM and
PDM codes. We selected 146 variable stars using criteria which focus
on long-period red variables (Miras and SRs), and investigated their
properties. Their observational data were tabulated.

In the ($\log P, \langle V_m \rangle$) and ($\log P, \langle R_m \rangle$)
diagrams, the luminosity of stars in the major cluster tended to
decrease as the period increases. In the ($\log P, \langle V_m \rangle -
\langle R_m \rangle$) diagram, the colours were redder as the
periods were longer.

We transformed $V_m, R_m$ magnitudes to $K_m$ values using 19
oxygen-rich stars, and presented the period-luminosity (PL) relation
in the ($\log P, K_m$) plane. Many of the stars in our sample,
including the large amplitude stars, fell near the position
corresponding to the O-Mira sequence by Feast et al. (1989) and Hughes
\& Wood (1990). A small number of stars (8) formed a separate group in
the period-magnitude or period-colour diagrams. This group (group $B$) was
found in the bright and short period region of the ($\log P, K_m$) plane
compared with the O-Mira sequence. These stars showed less periodic
variation, and six of them had small amplitudes.

   The separation in $\log P$ between them and the main group
   was $\sim$ 0.4.
   They were on the red side of the period-colour diagram and were
   located at the reddest and brightest tip of the colour-$K_m$ diagram
   similar to the long-period members.
   If our conversion scale is adequate, it indicates that some of the
   stars pulsate in a short period compared with the stars which may
   have a similar radius.

These facts were possibly explained by the assumption proposed by Wood
\& Sebo (1996): that those six stars with short periods and small amplitudes
were higher mode pulsators than the stars on the O-Mira sequence. 
Two particular stars, nlmc3-214266 and nlmc1-239264, were included
in group $B$. These stars showed different features from the other six
variable stars. Their pulsation property possibly differs from the
other six stars.

In the ($\log P, K_m$) diagram, below the O-Mira sequence, many stars
were located in a wide region. Since almost all identified C-Miras
were found in this domain, we supposed that a large number of
such variable stars were C-Miras which were more luminous than the
magnitude indicated here. In order to reduce the systematic scattering
in the ($\log P, K_m$) diagram, it was necessary to distinguish
carbon-rich stars from oxygen-rich stars in advance.

The difficulty in distinguishing oxygen-rich and carbon-rich stars comes from the fact that our blue filter covers both the standard $B$ and $V$ bands. It is expected that in the future these stars will be observed in the $B$ or $V$ bands which will distinguish between carbon-rich stars and oxygen-rich stars. Alternatively, using any passband which avoids the effect of TiO bands to estimate the bolometric magnitude of the stars could be used.  Improvement in the magnitude calibration and confidence in the interpretation of the plural sequences is expected to follow from the analysis of further large observational databases.

\section{Acknowledgments}

We are grateful to  A. Tanaka for his collaboration in preparing the PDM code.
We wish to thank P.A.Whitelock for her useful comments and help.
We also thank H. Saio for helpful discussions.
This work is supported by a grant-in-aid for scientific research (A)
of the Japan Ministry of Education, Science, Sports and Culture,
and also by the Marsden Fund of the Royal Society of New Zealand.

\label{lastpage}
\end{document}